\documentclass[a4paper, 11pt]{article}
\usepackage[colon]{natbib}
\usepackage{tabularx}
\usepackage{epsfig}
\usepackage[titletoc, title]{appendix}
\usepackage{setspace}
\usepackage{lscape}
\usepackage{booktabs}
\usepackage{array}
\usepackage{tabularx}
\usepackage[latin1]{inputenc}
\usepackage[english]{babel}
\usepackage{makeidx}
\usepackage{latexsym}
\usepackage{ifthen}
\usepackage{amssymb}
\usepackage{color,epsfig}
\usepackage{graphicx}
\usepackage{subfig}
\usepackage{mathrsfs}
\usepackage{wrapfig}
\usepackage{rotating}
\usepackage{natbib}
\usepackage{amsmath, amsthm}
\usepackage{nomencl}
\usepackage{float}
\usepackage{multirow}
\usepackage{multicol}
\usepackage{xcolor}
\usepackage{epic,eepic}
\usepackage{lscape}
\usepackage{tabularx}
\usepackage{pdflscape}
\usepackage{epstopdf}
\usepackage{calligra}
\usepackage{epsfig}
\usepackage[colon]{natbib}
\usepackage{tabularx}
\usepackage{amsmath}
\usepackage{amssymb}
\usepackage{pdflscape}
\usepackage{multirow}
\usepackage{multicol}
\usepackage[bottom]{footmisc}
\usepackage{array,arydshln}
\usepackage{appendix}
\usepackage{hyperref}

\newcolumntype{x}[1]{>{\centering\arraybackslash\hspace{0pt}}p{#1}}

\DeclareMathSizes {8}{8}{8}{8}
\makeindex           
\allowdisplaybreaks

\newcommand{\bs}{\boldsymbol}
\newcommand{\tbf}{\textbf}

\title{Quantile regression for longitudinal data:  unobserved heterogeneity and informative missingness}\label{cap6}

\author{Maria Francesca Marino\thanks{Dipartimento di Scienze Statistiche, Sapienza Universit\`a di Roma, Italy, {\tt mariafrancesca.marino@uniroma1.it}} \and Nikos Tzavidis \thanks{Department of Social Statistics \& Demography, University of Southampton, UK, {\tt n.tzavidis@soton.ac.uk}} \and Marco Alf\'o\thanks{Dipartimento di Scienze Statistiche, Sapienza Universit\`a di Roma, Italy, {\tt marco.alfo@uniroma1.it}}}

\date{}

\begin{document}

\maketitle
\onehalfspacing

\begin{abstract}
Linear quantile regression models provide a detailed and robust picture of the (conditional) response distribution as function of a set of observed covariates. Longitudinal data analysis is an interesting field of application of such models although this kind of data represent a substantial challenge, due to dependence issues. Often, dependence between measurements from the same units is modelled by considering sources of unobserved, individual-specific, heterogeneity. Quantile regression models have recently been extended to the analysis of longitudinal, continuous, responses, using time-constant, see e.g. \cite{GeraciBottai2007}, or time-varying, see \cite{Farcomeni2012}, random effects. In some empirical applications, however, we observe both temporal shocks in the overall trend and individual-specific heterogeneity in model parameters. To accommodate such situations, we propose to define a general quantile regression model for longitudinal, continuous, responses where time-varying and time-constant random parameters with unspecific distribution are jointly taken into account.
We further deal with the case of irretrievable, non ignorable, exit from the study (i.e. drop-out) and show how the proposed model can be interpreted in a pattern mixture perspective, where changes in the fixed effect vector are associated to mixture components describing the individual propensity to remain into the study. The proposed models are illustrated using a well known benchmark dataset on longitudinal dynamics of CD4 cells and a large scale simulation study.
\end{abstract}

\section{Introduction}\label{cap6_sec1}
In longitudinal studies, measurements recorded on the same individual are likely associated, and the adopted statistical modelling tools need to account for such a dependence. A common framework is based on postulating a conditional model which is augmented by individual-specific sources of unobserved heterogeneity that are meant to describe the dependence in the data. 
\\
In the context of longitudinal studies, \cite{GeraciBottai2007} have proposed a linear quantile model with random effects; this model has been extended, either from a structural or a computational perspective by \cite{LiuBottai2009} and \cite{GeraciBottai2013}. When random parameters are time-varying, the assumption of a time-constant distribution may lead to severe bias, see eg \cite{BartFarco2009} for a discussion. To solve this issue, \cite{Farcomeni2012} has proposed a linear quantile model with time-varying random intercepts, see e.g \cite{BartFarcoPennoni2012} for a general treatment of the topic. In some circumstances, it can be reasonable to assume that both time-constant and time-varying sources of unobserved heterogeneity influence longitudinal observations. In these cases, the proposal by \cite{Farcomeni2012} can be extended to deal also with (discrete) time-constant random parameters, inflating the number of states and forcing the transition probability matrix to be (quasi) diagonal at the cost of increase the computational complexity. To overcome this issue, we define a \emph{linear quantile mixed hidden Markov model}, where random parameters may be constant and/or varying over time. The model specification is not constrained to random intercepts only and, thus, offers greater flexibility;  modelling quantiles in place of the mean ensures robustness against the possible presence of outliers in the observed  data; accounting for structured, although unobserved, random variation leads to reliable and efficient parameter estimates. 
\\
A frequent feature of longitudinal studies is that some individuals may be unavailable at all the \textit{pre-determined} time occasions. The presence of missing information rises a number of challenges because of the potential bias in the parameter estimates. We describe how the model approach we propose can be interpreted in a pattern mixture perspective, according to \cite{Roy2003} and \cite{Roy2008}. With this representation, the dependence between the observed responses and the missing data process is due to the presence of sources of unobserved heterogeneity shared by individuals with a similar propensity to drop-out. 
This leads to groups characterized by common departures from the homogeneous linear quantile hidden Markov model and, therefore, to a very general approach. 
\\
The paper is structured as follows: in section \ref{lqhmm_model}, after discussing the proposal by \cite{Farcomeni2012}, we introduce the linear quantile mixed hidden Markov model, where time-varying and time constant random parameters are jointly considered. In section \ref{modellingLDO}, we show how the proposals by \cite{Roy2003} and \cite{Roy2008} can be extended to the quantile context and to time-varying random parameter models. The resulting model can be interpreted as a particular specification of the \emph{general} modelling framework we propose. The EM algorithm for parameter estimation is briefly sketched in section \ref{cap6_comp_details}. Further details are available in the supplementary material. Results of the application to the CD4 dataset \citep{Kaslow1987, ZegerDiggle1994} and of a large scale simulation study are given in sections \ref{cap6_simulations} and \ref{cap6_realData}, respectively. Further simulation-based evidence is included in the supplemntary material. Last section gives concluding remarks and outlines potential future research agenda.

\section{Linear quantile hidden Markov model}\label{lqhmm}
Let $Y_{it}$ be a continuous response variable and $\tbf x_{it}$ a set of covariates for unit $i = 1, ..., n$ recorded at time occasion $t = 1,\dots, T$. Quantile regression extends regression analysis for the centre of a conditional distribution; even when this represents the focus of analysis, median regression offers an outlier-robust alternative to least squares. {When analysing hierarchical (for example longitudinal) data, the dependence structure must be accounted for; a widely used approach is based on considering conditional models with random parameters; see \cite{LairdWare1982} for a general discussion.\\
In this framework, we rely on conditional independence given (individual-specific) latent characteristics (e.g. omitted covariates), and obtain marginal dependence due to measurements from the same individual sharing some common latent variables. Even when a (marginal) multivariate model does exist, the conditional approach may still be preferable, see \cite{Lee2004}.
\\
Random parameters enter in the model as random intercepts and/or random slopes and may be either  time-constant or time-varying.  When there is no or limited prior knowledge about the possible causes of unobserved heterogeneity, the time-varying option offers greater flexibility. \cite{Farcomeni2012} has proposed a linear quantile hidden Markov model (\emph{lqHMM}) that can be described as follows. For a given $\tau \in (0,1)$, let $\{S_{it}(\tau)\}$ be a homogeneous, first order, hidden Markov chain with state space $ \mathscr S(\tau) = \{ 1, ..., m(\tau) \}$; initial and transition probabilities are given by $\delta_{h}(\tau) =  \Pr(S_{it}(\tau) = h) $ and $q_{kh}(\tau) = \Pr(S_{it}(\tau) = h \mid S_{it-1}(\tau) = k), h,k = 1,\dots, m(\tau)$, respectively. The \textit{lqHMM} is defined by
\begin{equation}\label{cap4_farcoHMM}
Q(y_{it} \mid s_{it}, \tau) =  \tbf x_{it} ^\prime \bs \beta(\tau) +  \alpha_{s_{it}(\tau)} ,
\end{equation}
where $Q(\cdot \mid s_{it}, \tau)$ denotes the (conditional) quantile for the $i-$th individual, $i=1,\dots,n,$ being in state $s_{it}(\tau)$ at time $t$, and $\alpha_{s_{it}(\tau)}$ denotes an individual-specific random intercept that evolves over time according to the hidden Markov chain described before.
\\
This modelling structure can be enriched by considering individual-specific random slopes to account for individual departures from the fixed parameters $\boldsymbol \beta(\tau)$, thus relaxing the assumption of orthogonality between the observed covariates and the sources of unobserved heterogeneity. These random departures can be either time-constant, as in \cite{GeraciBottai2013}, or time-varying, as in a slighted extended version of \cite{Farcomeni2012}. In our perspective, a general model should account for both.

\section{Linear quantile mixed hidden Markov model}\label{lqhmm_model}
To define a general model specification, we assume that dependence is due to different sources of unobserved heterogeneity, some of which are time-varying, while others are time-constant. Let $\tbf b_i(\tau) = (b_{i1}(\tau),\dots, b_{iq}(\tau))$ be an individual-specific random parameter vector with density $f_{b} (\cdot \mid  \tbf D, \tau)$, where $\tbf D$ is a quantile-dependent covariance matrix, and $\text E(\tbf b_i(\tau)) = \tbf 0$. For a given $\tau$, we define a linear quantile mixed hidden Markov model (\emph{lqmHMM}) as follows
\begin{equation}\label{cap4_marinomHMM}
Q(y_{it} \mid s_{it}, \tbf b_i, \tau) = \tbf x_{it} ^\prime \bs \beta(\tau) + \tbf z_{it} ^\prime \tbf b_i(\tau) +  \tbf w_{it} ^\prime \bs \alpha_{s_{it}(\tau)}.
\end{equation}
In the expression above, $\bs \beta(\tau)$ denotes a $p$-dimensional vector of fixed parameters describing the effect of observed covariates on the $\tau$-th (conditional) response quantile, ${\bf z}_{it}$ and ${\bf w}_{it}$ are non-overlapping subsets of ${\bf x}_{it}$, while $\tbf b_i(\tau)$ and $\bs \alpha_{s_{it}(\tau)}$ identify time-constant and time-varying random deviations from $\bs \beta(\tau)$, respectively.
%
All model parameters introduced so far are indexed by  $\tau$; in what follows, we simplify the notation by dropping the $\tau$ index.
\\
The \textit{lqmHMM} is based on the following assumption.
The random vector $\tbf b_i$ and the hidden Markov process $\{S_{it}\}$ are assumed to be independent as they are meant to capture different sources of unobserved heterogeneity. The distribution of the response variable is defined conditional on the hidden state occupied at time occasion $t$, i.e. $s_{it}$, and the (time-constant) individual-specific random vector $\textbf{b}_i$; conditional on these parameters, longitudinal observation from the same individual are independent (local independence assumption) and the conditional distribution of the individual sequence is given by:
\begin{equation}\label{cap5_cond_individualSeq}
f_{y \mid s,b} (\tbf y_i \mid \tbf s_i, \tbf b_i, \bs \psi,\tau) = \prod_{t = 1}^{T} f_{y \mid s,b} (y_{it} \mid y_{i1:t-1}, s_{i1:t}, \tbf b_i, \tau) = 
\prod_{t = 1}^{T} f_{y \mid s,b} (y_{it} \mid s_{it}, \tbf b_i, \tau),
\end{equation}
where $\bs \psi = (\bs \beta, \bs \alpha_1, \dots, \bs \alpha_m)$ is the vector of longitudinal model parameters.\\
In expression (\ref{cap5_cond_individualSeq}), $y_{i1:t-1}$ denotes the response history for the $i$-th individual up to time $t-1$ and $s_{i1:t}$ is the sequence of hidden states up to time $t$. It is worth to notice that the model in equation (\ref{cap4_marinomHMM}) reduces to the model by \cite{GeraciBottai2007} when a single state is considered ($m=1$) and to the model by \cite{Farcomeni2012} when $ w_{it} =  1$ and $b_i =  0$, $\forall i=1, \dots, n, t = 1, \dots, T_i$. 
In order to derive parameter estimates in a maximum likelihood perspective, we adopt an asymmetric Laplace distribution \citep{Yu2005} for the longitudinal responses, as in \citealt{GeraciBottai2007}; that is
\[
Y_{it} \mid s_{it}, \tbf b_i \sim \text{ALD} \left (\mu_{it}[s_{it}, \tbf b_i], \sigma, \tau \right),
\]
where the location parameter $\mu_{it}$ is defined by expression \eqref{cap4_marinomHMM}. Based on this assumption, expression \eqref{cap5_cond_individualSeq} can be written as
\begin{equation}\label{indseqdensity}
f_{y \mid s, b} (\tbf y_i \mid \tbf s_i, \tbf b_i, \bs \psi, \tau) = \left[\frac{\tau (1-\tau)}{ \sigma}\right] ^{T_i}
\exp \left\{ -\sum _{t = 1}^{T_i}\rho_{\tau} \left[ \frac{ y_{it} - \mu_{it} [s_{it}, \tbf b_i]
}
{\sigma} \right] \right\},
%
\end{equation}
with $\rho_\tau(\cdot)$ denoting the quantile asymmetric loss function  \citep{KoenkerBassett1978}. 
\\
Let $\bs \Phi = (\bs \psi, \bs \delta, \tbf Q, \tbf D) $ be the vector of  all model parameters; the observed data likelihood is defined by 
\begin{equation}\label{cap4_likIntegral}
L(\bs \Phi \mid \tbf y, \tau) = \prod_{i = 1}^n \int \left\{\sum_{\tbf s_i} \left[
\prod_{t = 1}^{T_i} f_{y \mid s,b} (y_{it} \mid s_{it}, \tbf b_i, \bs \psi, \tau) \right]
f_s \left( \textbf s_i \mid \boldsymbol \delta, \textbf Q, \tau\right)\right\}
f_b (\tbf b_i \mid \tbf D, \tau) 
\textrm d \tbf b_i,
\end{equation}
where, due to the Markov property, the marginal distribution for the hidden chain can be factorized as
\begin{equation*}
f_s \left( \textbf s_i \mid \boldsymbol \delta, \textbf Q, \tau\right) = 
f_s \left (s_{i1} \mid \bs \delta, \tau \right) \prod _{t = 2} ^ {T_i}  f_s \left(s_{it} \mid s_{it-1}, \tbf Q, \tau \right) = \delta_{s_{i1}} \prod_{t = 2}^{T_i} q_{s_{it-1}s_{it}}.
\end{equation*}
\\
For a parametric specification of $f_b (\cdot \mid \tbf D, \tau)$, parameter estimation would require the solution of a multiple integral that does not have a closed form solution. The next section entails the choice of such a distribution and its effect on the likelihood approximation.

\subsection {Estimating the random parameter distribution}
\label{sec_NPML}
When the random parameter distribution is parametrically specified, one may use a Monte Carlo EM algorithm, as in \cite{GeraciBottai2007} and \cite{LiuBottai2009}, or a direct ML approach with Gaussian quadrature, as in \cite{GeraciBottai2013}. Both should be appropriately extended to deal with the hidden Markov chain. When a limited number of repeated measurements are available for each individual, the choice of the random parameter distribution can be crucial. To avoid  potential misspecification, we propose to approximate it by using a discrete distribution on $G(\tau) \le n$ support points $\tbf b_g$ with masses $\pi_g = \Pr(\tbf b_g)$, $\pi_{g} \ge 0, \sum_g \pi_g = 1, g = 1, ..., G$. 
This is known as the nonparametric maximum likelihood (NPML) estimate of the mixing distribution $f_b(\cdot \mid {\bf D, \tau})$, see \cite{Aitkin1999}, \cite{Bohning1982} and \cite{Lindsay1983a, Lindsay1983b}; locations and masses are treated as unknown parameters and estimated through the observed data. Let us introduce a discrete latent variable $\boldsymbol \zeta_i = (\zeta_{i1},\dots, \zeta_{iG})$, to represent component membership. That is, $\zeta_{ig}=1$, $g = 1,\dots, G$, if the $i$-th individual belongs to the $g$-th component of the mixture, zero otherwise. As before, $\boldsymbol \zeta_i$ and $G$ depend on the chosen quantile $\tau$, but we will suppress this indexing to simplify the notation. Denoting by $\bs \psi = (\bs \beta, \tbf b_1, \dots, \tbf b_G, \bs \alpha_1, \dots, \bs \alpha_m)$ the set of longitudinal model parameters and by $\bs \pi = (\pi_1, \dots, \pi_G)$ the vector of mixture component probabilities, the observed data likelihood in equation \eqref{cap4_likIntegral} becomes 
\begin{equation}\label{cap4_likNPML}
L(\bs \Phi \mid \tbf y, \tau) = \prod_{i = 1}^n \sum_{g = 1}^G \left\{\sum_{\tbf s_i}\left[ 
\prod_{t = 1}^{T_i} f_{y \mid s,\zeta} (y_{it} \mid s_{it}, \zeta_{ig}=1, \bs \psi, \tau) \right] f_s \left( \textbf s_i \mid \boldsymbol \delta, \textbf Q, \tau\right)
\right\} \pi_{g}, 
\end{equation}
where $\bs \Phi = (\bs \psi, \bs \delta, \tbf Q, \bs \pi)$. 
The computational complexity of the proposed approach is linear with the integral dimension, while it increases exponentially in the Gaussian quadrature approach, see \cite{MaruottiRyden2009}. Since locations in the finite mixture are completely free to vary over the corresponding support, extreme departures from the homogeneous model can be accommodated. Direct maximization of the likelihood in equation \eqref{cap4_likNPML}, although possible, can be challenging. A generalization of the EM algorithm for finite mixtures provided by \cite{Aitkin1999} offers a simple alternative. In Section \ref{cap6_comp_details} we briefly sketch the structure of the algorithm. Further detail can be found in the supplementary material.

\section{Handling non-ignorable drop-out}\label{modellingLDO}

In this Section we extend the \textit{lqmHMM} formulation to handle potentially non-ignorable drop-out. Drop-out is a common problem in longitudinal data since individuals may leave the study before its planned end. While unbalanced designs do not pose particular problems, the question is whether missing data may bias parameter estimates.
Let $\tbf R_i = (R_{i1}, ..., R_{iT})$ denote the missing data indicator vector for the $i$-th individual, where $R_{it} = 1$ if $y_{it}$ has not been observed at time $t = 1, ..., T$ and zero otherwise. Since we are considering drop-out, that is irretrievable exit from the study, $R_{it} = 1 \Longrightarrow R_{it^{\prime}} = 1, t^\prime \geq t = 1, ..., T$. \\
Let $\bs \Phi$ and $\bs \gamma$ denote the parameter set for the longitudinal and the missing data model, respectively. \cite{Little1993} and \cite{LittleRubin2002} define two broad classes of models to handle (potentially) non-ignorable missing data. 
In the selection model (SM) formulation, see e.g. \cite{Heckman1976}, the joint distribution of $\tbf y_i$ and $\tbf r_i$ can be factorized as 
\begin{equation}\label{cap6_selection}
f_{y, r}(\tbf r_i, \tbf y_i \mid \bs \Phi, \bs \gamma) = f_{r \mid y}(\tbf r_i \mid \tbf y_i, \bs \gamma) f_{y}(\tbf y_i \mid \bs \Phi), \quad i = 1, ..., n,
\end{equation}
where the conditional density $[\tbf r_i \mid \tbf y_i, \bs \gamma]$ describes the selection mechanism leading each unit to continue or stop participating in the study. In the pattern-mixture model (PMM) formulation, the following factorization holds
\begin{equation}\label{cap6_sec_PMM}
f_{y, r}(\tbf r_i, \tbf y_i \mid \bs \Phi, \bs \gamma) = f_{y \mid r}(\tbf y_i \mid \tbf r_i, \bs \Phi ) f_{r}(\tbf r_i \mid \bs \gamma), \quad i = 1, ..., n.
\end{equation}
The rationale for PMMs is that each subject has its own propensity to drop-out from the study; individuals dropping-out from the study closer in time have similar propensities and share some common observed and/or unobserved features. Therefore, the model for the whole population is given by a mixture over these patterns. Further modelling alternatives (e.g. shared parameter models -- SPMs) are reviewed by \cite{Little1995} and \cite{RizopoulosLesaffre2014}. In the hidden Markov literature, \cite{Bartolucci2015} discuss a model for multivariate longitudinal responses and a (discrete) time to event, with discrete time-varying and time-constant random intercepts shared by the longitudinal response and the missingness indicator. A pattern mixture approach has been described by \cite{Maruotti2015}, where the transition matrix is a function of the number of available measurements for the individual. 
\\
According to this latter proposal, we define a PMM to account for the potential presence of informative missingness; we assume to be interested in the conditional quantiles of the response variable and adopt a lqHMM formulation. 
To overcome the weak identifiability of PMMs due to a possibly large number of patterns \citep[see e.g.][]{Roy2003}, we consider a reduced number of classes, representing different propensities to drop-out from the study; in the following, we will refer to them as latent drop-out (LDO) classes.\\
Let $T_i = T - \sum_{t = 1}^T R_{it}$ denote the number of measurements available for the $i$-th individual, $i=1,\dots,n$. The latent variable $\boldsymbol \zeta_i = (\zeta_{i1},\dots, \zeta_{iG})$ which is used in the \textit{lqmHMM} to denote component membership, is now used to denote the membership to a specific LDO class. Sample units in the same LDO class share some common latent characteristics that lead to changes in the covariate effect estimates. According to \cite{Roy2003} and \cite{Roy2008}, we assume that individuals with a higher propensity to remain into the study have a higher chance to present complete responses, i.e. they have higher values for $T_{i}$. 
Hence, the probability of being in one of the first LDO classes is described by a monotone function of the number of available measurements $T_{i}$. For a given $\tau \in (0,1)$, the following (proportional) ordinal regression model is used
\begin{equation}\label{cap6_royModel_ldo}
\Pr \bigg(\sum_{l= 1}^g \zeta_{il} = 1 \mid T_i, \tau \bigg) = \frac{\exp \left(\lambda_{0g} + \lambda_{1}T_{i}\right)}{1+ \exp\left(\lambda_{0g} + \lambda_{1}T_{i}\right)}
\end{equation}
under the constraints $\lambda_{01} \leq \dots \leq \lambda_{0G-1}$. Two issues should be noticed. First, by using a LDO-based approach, we often consider a reduced number of classes, and this help solve the weak identifiability issues common to PMMs. Second, this model specification generalizes the \textit{lqmHMM}, since the latent variable $\bs \zeta_i$ is now ordinal and the corresponding masses are defined to be a function of $T_i$. We assume that, conditional on $S_{it} = s_{it}$ and $\zeta_{ig} = 1$, longitudinal observations from the same individual are independent (local independence assumption); further, $\bs \zeta_i$ capture all the dependence between the longitudinal and missing data process; conditional on such latent variables, the two processes are no longer dependent. As before, longitudinal responses follow a conditional ALD, with location parameter defined by 
\begin{equation}\label{cap6_lqHMM_LDO}
Q_{\tau} (y_{it} \mid s_{it}, \zeta_{ig}  =1, \tau) = \tbf x_{it}^\prime \bs \beta + \tbf z_{it} ^\prime\tbf  b_g + \tbf w_{it} ^\prime \bs \alpha_{s_{it}},
\end{equation}
where the dependence of model parameters on $\tau$ has been suppressed for ease of notation.
Denoting by $\tbf y_i^o$ and $\tbf y_i^m$ the observed and the missing part of the individual sequence $\tbf y_i$, the observed data likelihood for the $i$-th individual is given by
\begin{align}
L_{(i)}(\bs \Phi, \bs \gamma \mid \tbf y_i^o, T_i, \tau) & = 
\sum_{\tbf s_i}\sum_{ g = 1}^G \int_{\tbf y_i^m} 
f_{y \mid s \, \zeta} (\tbf y_{i} \mid \tbf s_{i}, \zeta_{ig} = 1, \bs \psi, \tau)
 \, f_s(\tbf s_i \mid \bs \delta, \tbf Q, \tau) 
 \times \nonumber \\ 
&   \quad \quad
\times   
\pi_{ig} (T_i \mid\bs \lambda, \tau)
f_T(T_i \mid \bs \gamma, \tau) \mathrm d \tbf y_i ^m,
\end{align}
where $\pi_{ig} (T_i \mid \bs \lambda, \tau) = f_{\zeta \mid T} (\zeta_{ig}  =1  \mid T_i, \bs \lambda, \tau)$ is the conditional probability for the $i$-th individual in the $g$-th LDO class, $g = 1, ...,G$, and is defined as the difference between two adjacent cumulative logits, see \cite{Agresti2010}. By assuming that the parameters $\bs \Phi$ and $\bs \gamma$ are separate, the marginal distribution of the missing data process $f_T (T_i \mid \bs \gamma, \tau)$ can be left unspecified and inference can be based on the conditional observed data likelihood
\begin{align}\label{cap6_cond_obsLik}
L_{(i)}(\bs \Phi\mid  \tbf y_i^o, T_i, \tau) &=  \sum_{\tbf s_i} \sum_{g = 1}^G
f_{y \mid \, s \zeta} (\tbf y_{i}^o \mid \tbf s_{i}, \zeta_{ig}  =1, \bs \psi, \tau) \, f_s(\tbf s_i \mid \bs \delta, \tbf Q, \tau)  
 \pi_{ig} (T_i \mid \bs \lambda, \tau).
\end{align}
We may notice that expression \eqref{cap6_cond_obsLik} reduces to \eqref{cap4_likNPML} when $\pi_{ig} (T_i \mid \bs \lambda, \tau) = \pi_{g}$, $\forall i=1,\dots,n$ and $g=1,\dots,G$, that is when  $\lambda_{1}=0$ in equation \eqref{cap6_royModel_ldo}. Hence, mixture components are introduced either as a way to approximate intractable integrals (as in \textit{lqmHMM}) or to describe time-constant unobserved characteristics that are related to the drop-out mechanism (as in \textit{lqHMM+LDO}).

\section{ML estimation} \label{cap6_comp_details}
Parameter estimates for the \textit{lqmHMM} and the \textit{lqHMM+LDO} are obtained by using the Baum-Welch algorithm \citep{McDonaldZucchini1997}. In this section, we briefly sketch the algorithm; a more detailed discussion is available in the Supplementary material. As before, we suppress the $\tau$ indexing of the model parameters. We will refer to LDO classes with the generic term ``components'', to align the terminology of \textit{lqmHMM} and \textit{lqHMM+LDO}, and use $\pi_{ig}$ in place of $\pi_{ig} (T_i \mid \bs \lambda, \tau)$ for the \textit{lqHMM+LDO} formulation, while $\pi_{ig} = \pi_{g}, \forall i = 1, \dots, n$ for the \textit{lqmHMM} one.
Let $u_{i}(h) = \mathbb I \left[S_{it} = h \right]$ denote the indicator variable for the $i$-th individual being in the $h$-th state at time occasion $t$ and $u_{it}(k,h) = \mathbb I \left[ S_{it-1} = k, S_{it} = h \right]$ indicating whether he/she moves from the $k$-th state at time occasion $t-1$ to the $h$-th one at time occasion $t$. 
As before, $\zeta_{ig}$ is the indicator variable for the $i$-th unit belonging to the $g$-th component. Starting from the definition of the conditional complete data log-likelihood
\begin{align} \label{auxiliary_NPML}
 \ell_c(\bs \Phi \mid \tbf y, \tbf T, \tbf S, \bs \zeta,\tau) &  \propto
\sum _{i = 1} ^ n \bigg\{
\sum_{h = 1}^m u_{i1}(h) \log \delta_{h} +
\sum_{t=2}^{T_i} \sum_{h,k=1}^m u_{it}(k,h) \log q_{kh} + 
\sum_{g = 1}^G \zeta_{ig} \log \pi_{ig} +
\nonumber \\[0.1cm]
& - T_i \log (\sigma) - 
\sum _{t = 1}^{T_i}  \sum_{h=1}^m  
\sum_{g = 1}^G u_{it}(h) \zeta_{ig}
\rho_{\tau} \left[ 
		\frac{ y_{it} - \mu_{it}[S_{it} = h, \tbf b_g] }{\sigma} 
\right]  
 \bigg\},
\end{align}
parameter estimates are derived by using an EM-type algorithm, that is by alternating an E- and a M-steps. 
In the E-step we take the expected value of the complete data loglikelihood \eqref{auxiliary_NPML}, given the observed data and the current parameter estimates; we refer to this as $Q(\bs \Phi \mid \bs \Phi^{(r)})$. This amounts to replacing the indicator variables by the corresponding (posterior) expected values. Once these quantities have been computed, in the M-step, model parameter estimates are derived by maximizing $Q(\bs \Phi \mid \bs \Phi^{(r)})$. Given the separability of the parameter spaces for the longitudinal and the missing data process, the maximization can be partitioned into independent sub-problems, which  considerably simplifies the computations. See the Supplementary Material for details.\\
The E- and the M-step of the algorithm are iterated until convergence, that is until the (relative) difference between subsequent likelihood values is lower than an arbitrary small amount $\varepsilon$. Penalized likelihood criteria, such as BIC \citep{Schwarz1978}, can be used to jointly identify the best number of components and hidden states. As it typically happens in the linear quantile mixed model framework, standard errors for parameter estimates are obtained by nonparametric block bootstrap.
This amount to resampling individual indexes and keep all the corresponding measurements in order to preserve the within individual dependence structure, see e.g. \cite{Lahiri1999} for references.

\section{Application: Re-analysing the CD4 cell count data}\label{cap6_realData}
\subsection{Data Description}\label{cap6_Datades}
The models we propose are illustrated by re-analysing the CD4 cell count dataset \citep{ZegerDiggle1994}. Data come from the Multicenter AIDS Cohort Study (MACS), started in 1984 and involving more than $5000$ volunteered gay and bisexual men from Baltimore, Pittsburgh, Chicago and Los Angeles. 
The HIV virus destroys the T-lymphocytes (CD4 cells) which play a vital role in immune function; the virus progression can therefore be assessed by measuring the number of CD4 cells, which, on average, tend to decrease throughout the incubation period. Among the volunteers participating in the study, $371$ ($7\%$) seroconverted during the analysed time window. Since the aim was to understand the impact of serconversion of serconversion on the dynamics of the CD4 count, we have considered in our analysis, coherently with \cite{ZegerDiggle1994}, $2376$ measurements from $369$ individuals (two of them have been discarded due to missing values in the covariates), observed from a minimum of $3$ years before to a maximum of $6$ years after the seroconversion. The observed data suffer from attrition and, for each individual, the number of available measurements ranges from  a minimum of $1$ to a maximum of $12$.
%
\\
The interest is at determining the effect of a set of covariates on the evolution of the CD4 cell counts over time while controlling for sources of unobserved heterogeneity. The set of covariates includes: years since seroconversion (negative values indicate the current CD4 measurement was taken before the seroconversion), age at seroconversion (centered around $30$), smoking (packs per day), recreational drug use (yes or no), number of sexual partners, depression symptoms  measured by the CES-D scale \citep{Radloff1977}, ranging from $0$ to $60$, with larger values indicating more severe symptoms. The analysis was conducted on the log transformed CD4 counts, that is $\log (1+\text{CD4 count})$.
We start the analysis by assuming a non-informative drop-out scenario, and compare the obtained results with those from the corresponding non-ignorable missing data model, the \textit{lqHMM+LDO}.

\subsection{Linear quantile mixed hidden Markov model}\label{cap5_application}
To model the evolution of individual trajectories over time, we have considered individual-specific, time-varying and time-constant, random parameters.
From a preliminary exploratory analysis, and by looking at the fit of a number of regression models, we have decided to focus on the following model specification: 
\[
Q_\tau(y_{it} \mid s_{it}, \tbf b_i) = 
\tbf x_{it} ^\prime \bs \beta(\tau) + \tbf z_{it} ^\prime \tbf b_i(\tau) +  \tbf w_{it} ^\prime \bs \alpha_{s_{it}(\tau)},
\]
where, $\tbf x_{it}$ includes two continuous covariates (time since seroconversion and age), the dummy variable drug (baseline: no) and three discrete variables (packs of cigarette per day, number of sexual partners and CES-D score). The vectors associated with the time-constant and the time-varying random parameters, $\tbf z_{it}$ and $\tbf w_{it} = \tbf w_i$, include the time since seroconversion and a column of ones, respectively. That is, the model considers a set of fixed parameters, a time-varying random intercept and a time-constant random slope.
\\
We have fit the proposed model for a varying number of hidden states and mixture components and for three quantiles, $\tau = (0.25, 0.50, 0.75)$.  To reduce the chance of being trapped in local maxima, we have adopted a multi-start strategy, which can be described as follows. Deterministic starting solutions for the initial and the transition probabilities have been obtained by setting $\delta_h = 1/m$ and $q_{kh} = (1 + s \mathbb I(h = k)) /(m + s)$, for a suitable constant $s>0$.
Starting values for fixed parameters have been obtained by fitting a standard linear quantile regression model,
 while for the time-varying and the time-constant random parameters, $m$ and $G$ Gaussian quadrature locations are added to the corresponding fixed effects. To obtain a set of random starting values, we have randomly perturbed these deterministic starts. For each combination of the number of hidden states and the number of mixture components, we have considered $30$ random start, and retained the best solution according to the BIC, see Table \ref{cap5_tab_CD4_BIC}.

\begin{center}
Table \ref{cap5_tab_CD4_BIC} about here
\end{center}

The results suggest to select a model with $m=4$ hidden states at all the analysed quantiles, thus highlighting quite a strong time-varying unobserved heterogeneity. The distribution for the individual-specific slope associated to $\text{Time}_{\text{sero}}$ is approximated by a discrete distribution with a number of mixture components that decreases as we move towards the right tail of the response distribution. As it can be evinced by looking at Table \ref{cap5_tab_CD4_BIC}, we choose $G = 5,4,3$ components for $\tau = 0.25$, $\tau =  0.50$ and $\tau = 0.75$, respectively. In Table \ref{cap5_tab_CD4_estimLong}, we report the parameter estimates for the longitudinal data model, with $95\%$ confidence intervals based on $B = 1000$ block bootstrap samples. 

\begin{center}
Table \ref{cap5_tab_CD4_estimLong} about here
\end{center}

As it can be noticed by looking at Table \ref{cap5_tab_CD4_estimLong}, for all the analysed quantiles, age appears to play a minor role while the packs of cigarettes and the number of sexual partners have a positive and significant effect on the log count of CD4 cells. At the tails of the response distribution, the recreational use of drugs has a positive effect while this is negligible at the median as it is clear by looking at the corresponding confidence intervals. According to \cite{ZegerDiggle1994} the positive effect associated to some ``risk'' factors reflect a selection bias mechanism, where healthier men stay longer in the study and choose to continue their usual practices. More severe depression symptoms, indicated by higher values of the CES-D score, imply a slight decrease in the number of T-lymphocytes. For the time since seroconversion, the effect is negative; that is, the number of CD4 cells decreases with increasing lag from the date of seroconversion (see the $\text{Time}_\text{sero}$ effect in Table \ref{cap5_tab_CD4_estimLong}). This effect reduces when we move towards higher quantiles, suggesting that the progression in time of the virus is slower for healthier men. The estimated variance of the random parameters ($\sigma_b$) is significantly different from zero and reveals the presence of substantial individual-specific departures from the homogeneous effect of $\text{Time}_\text{sero}$ on the T-lymphocyte counts; as for the number of components, also this variability reduces when moving towards the right tail of the response distribution. We may also notice that, when we move from $\tau = 0.25$ to $\tau = 0.75$, state-dependent intercept estimates tend to increase, and this is coherent with increasing values of baseline (log) CD4 levels. Table \ref{cap5_tab_CD4_estimQ} reports the estimated initial and transition probabilities for the hidden Markov chain. The combination of these results with the intercept values reported in Table \ref{cap5_tab_CD4_estimLong} helps us understand the evolution of CD4 cell counts over time, conditional on the observed covariates.  

\begin{center}
Table \ref{cap5_tab_CD4_estimQ} about here
\end{center}

For all the analysed quantiles, the  estimated initial probabilities suggest that most of the sample units start the study with intermediate levels of CD4 cell counts ($\delta_2 + \delta_3 > 0.70$), while few observations start with more extreme (lower or higher) levels. For $\tau = 0.50$ and $\tau=0.75$, transitions between hidden states over time are quite unlikely ($q_{hh} >0.8$) and, if any transition is observed, units tend to move towards states with lower values of the intercept. For $\tau = 0.25$, we observe a slightly different evolution of the response over time. Estimated transition probabilities highlight that, for less healthy men, the number of CD4 cells in the blood tends to repeatedly increase and decrease over the follow-up time, and this is particularly evident for ``lower'' hidden states. Transitions towards the first hidden state (with the lower CD4 log-count) are unlikely $(\sum_{k = 1}^m q_{k1}<0.15)$ and, if any transition to the first hidden state is observed, in the next occasion individuals tend to move towards states characterized by higher CD4 cell count levels ($q_{11} = 0.284$), that is, the sudden decrease to the fist hidden state is just temporary.

\subsection{Linear quantile hidden Markov model with latent drop-out classes}\label{cap6_application}
As we have already discussed, each individual has been observed from a minimum of $3$ years before to $12$ years after seroconversion.
Table \ref{cap6_tab_missing} shows the number of individuals remaining in the study at each measurement occasion. Only a small portion of individuals has been observed until the end of the follow-up time. 

\begin{center}
Table \ref{cap6_tab_missing} about here
\end{center}
We report in Figure \ref{cap6_fig_respDist_vs_TDO} the distribution of the response variable at each time occasion, stratified by whether or not units drop-out from the study between the current and the subsequent measurement; as it is clear, CD4 levels tend to suddenly decrease just before the units drop-out of the study, especially in the first measurements occasions.  

%

\begin{center}
Figure \ref{cap6_fig_respDist_vs_TDO} about here
\end{center}

These findings suggest that healthier individuals tend to stay longer into the study and that a potential dependence between the longitudinal and the missing data process may be present. 
As we have already highlighted, individual-specific heterogeneity in the slope for $\text{Time}_\text{sero}$ decreases when moving from the first to the third quartile (that is when moving from sicker to healthier men); here, we aim at analysing if such changes may somehow be related to the missing data process. For this purpose, wee have defined the following \emph{lqHMM+LDO}

\[
Q_\tau(y_{it} \mid s_{it}, \zeta_{ig} = 1) = 
\tbf x_{it} ^\prime \bs \beta(\tau) + \tbf z_{it} ^\prime \tbf b_g(\tau) +  \tbf w_{it} ^\prime \bs \alpha_{s_{it}(\tau)},
\]
to compare the results with those we have obtained by the MAR counterpart, the \emph{lqmHMM} we have discussed before. The vectors $\tbf x_{it}$, $\tbf z_{it}$ and $\tbf w_{it}$ are defined as in the \textit{lqmHMM} specification and parameters have all the same interpretation but the random slope for $\text{Time}_\text{sero}$, which is now assumed to vary across LDO classes. We have fit the proposed model for $\tau=(0.25, 0.50, 0.75)$. To avoid local maxima, model parameters have been initialized via the same multi-start strategy we have described for the \emph{lqmHMM}. Initial values of the missing data model parameters have been obtained by fitting an ordered logit model on the discretized times to drop-out, randomly perturbed to avoid infinite estimates for the $T_i$ effect.
For each combination of the number of hidden states and LDO classes, we have considered $30$ starting points and retained the solution with the best BIC value. Results are reported in Table \ref{cap6_tab_CD4_BIC}. 
\begin{center}
Table \ref{cap6_tab_CD4_BIC} about here
\end{center}
We select the model with $m = 5$ hidden states and  $G = 5$ LDO classes when modelling the first quartile of the response distribution  ($\tau = 0.25$); for the median and the third quartile, the solution with $m = 4$ and $G = 4$ provides the lowest BIC values. By looking at the parameter estimates for the LDO class model at $\tau = 0.75$, we noticed that two $\hat{\lambda}_{0g}$ estimates do not significantly differ from zero and the corresponding confidence intervals overlap. Therefore, to avoid spurious solutions, we have decided to search for the optimal number of classes within the set of models with $G \leq 3$. As a result, for $\tau = 0.75$, the best fit corresponds to $m = 4$ hidden states and $G = 3$ LDO classes. In Table \ref{cap6_tab_long}, we report the estimated parameters (state-dependent intercepts and fixed slopes) for the longitudinal data model with corresponding $95\%$ confidence intervals, in parentheses, based on  $B = 1000$ block bootstrap samples. 
\begin{center}
Table \ref{cap6_tab_long} about here
\end{center}
If we compare results in Table \ref{cap6_tab_long} with those obtained by the \emph{lqmHMM} and reported in Table \ref{cap5_tab_CD4_estimLong}, we may observe only slight changes in the fixed parameter estimates for Packs and $\text{Time}_\text{sero}$ at $\tau=0.25$, while all other fixed parameters seem unchanged. 
As expected, estimates of the state-dependent intercept increase when moving from the left to the right tail of the response distribution. By combining these results with the estimated initial and transition probabilities reported in Table \ref{cap6_tab_HMM}, we notice that we have obtained findings that are similar to those we have discussed for the \textit{lqmHMM} specification. Only for $\tau=0.25$ we observe a further state with a lower intercept estimate that seem to be linked to highly variable dynamics of the response for units dropping-out early from the study; for $\tau= \{0.50, 0.75\}$, differences seem to be negligible.
\begin{center}
Table \ref{cap6_tab_HMM} about here
\end{center}
In Table \ref{cap6_tab_ldoLongPar}, we show the estimated LDO-dependent locations side by side with the location estimates for the \emph{lqmHMM} specification. 
\begin{center}
Table \ref{cap6_tab_ldoLongPar} about here
\end{center}
In both models, the effect of $\text{Time}_{\text{sero}}$ on the CD4 cell count is negative, significant and the estimate reduces when moving from $\tau = 0.25$ to $\tau = 0.75$. Location estimates suggest that individuals belonging to ``lower'' mixture/LDO components have a steep reduction in the (log) number of T-lymphocytes when the time since seroconversion increases.
%
 This effect progressively reduces for units belonging to ``higher'' latent categories. The results obtained from the \emph{lqHMM+LDO} can be further explored by looking at the LDO class model estimates reported in Table \ref{cap6_tab_ldoModel}. 
For all the analysed quantiles, the negative effect of the time to drop-out ($\hat \lambda_1<0$) suggests that ``lower'' LDO classes identify groups of individuals with shorter longitudinal sequences: the probability of belonging to one of the first $g$ classes reduces when the number of available measures increases. 
\begin{center}
Table \ref{cap6_tab_ldoModel} about here
\end{center}
%

These results are clearer when looking at Figure \ref{fig_spaghetti_unique}, where we report the longitudinal trajectories of individuals classified into the different LDO components under the \textit{lqHMM+LDO} formulation via a MAP rule for $\tau = (0.25, 0.50, 0.75)$. 
Local polynomial regression curves (blue lines), $95 \%$ confidence intervals (gray bands) and mean values (blue dots) are reported to highlight the general trend. Due to the missing data process, wider confidence intervals are observed for the last measurement occasions. 
As it is clear, higher LDO classes correspond to longer longitudinal sequences. 
\\ 
We use different colors to highlight those individuals that, under the \textit{lqmHMM} formulation, have not been classified in the same component as in the \textit{lqHMM+LDO} one. Red and yellow trajectories identify individuals that have been moved forward (higher classes, red) and backward (lower classes, yellow), respectively, when estimating the \textit{lqHMM+LDO}. Black trajectories represent the evolution over time of the CD4 count levels for those individuals that have been classified likewise by the two model specifications. 

\begin{center}
Figure \ref{fig_spaghetti_unique} about here
\end{center}

By looking at 
these figures, it is clear that, generally, the two models offers quite  a similar classification when we look at the component membership for $ \tau = 0.50$ and $\tau = 0.75$ are concerned. As we have highlighted before, the effect of the missing data process on the longitudinal response is quite negligible when considering individuals in better health conditions, even if some anomalies still seem to be corrected when modelling the missing data process: some individuals with longer (respectively shorter) sequences and weaker (respectively stronger) reduction of the response values over time are moved in higher (respectively lower) LDO classes under the lqHMM+LDO formulation.
\\
As regards $\tau = 0.25$ (i.e. for responses associated with less healthy individuals, which often represent the main target of inference), classifications supplied by the two modelling approaches seem to be quite different, thus confirming the stronger impact of the missing data process on the first quartile of the response distribution. Individuals are mainly shifted in lower LDO classes when compared to \textit{lqmHMM} results. These classes are characterized both by shorter longitudinal sequences and by a stronger impact (especially in the last observed occasions) of $\text{Time}_\text{sero}$ on the CD4 count levels (also stronger than those identified by \textit{lqmHMM}) which seem to be coherent with the individual path shown in the first panel of Figure \ref{fig_spaghetti_unique}.
These results, together with the lower BIC values obtained under \emph{lqHMM+LDO} (see Tables \ref{cap5_tab_CD4_BIC}-\ref{cap6_tab_CD4_BIC}), suggest a better fit and, thus, render \emph{lqHMM+LDO} an interesting modelling solution for the analysis of such kind of data, especially when we look at the first quartile of the response distribution. The observed increase in the log-likelihood values we obtain when moving from \emph{lqmHMM} to \emph{lqHMM+LDO} could be ascribed to a more flexible structure for the component priors; in the \emph{lqmHMM} these are constant across individuals while, in the \emph{lqHMM+LDO}, they depend on individual-specific features. In our formulation, we assume that such features are connected to a differential propensity to stay in the study, which, in turn, is summarized by $T_i$. Nevertheless, this propensity is unobservable and, therefore, we may not conclude that the missing data process is truly informative, since $T_i$ could represent other, unobserved, individual characteristics that are not linked to the propensity to drop-out. A \textcolor{blue}{sensitivity analysis} to check for non-ignorability of the missing data generating process represents a further step to validate the model.

\section{Simulation study}\label{cap6_simulations}
To study the performance of the proposed models, we have implemented a large scale simulation study made up by two different scenarios. 
First, to evaluate the empirical behaviour of \textit{lqmHMM}, we have considered a scenario (Scenario 1) with completely observed longitudinal responses and compared our proposal with the lqHMM specification we obtain when setting $G=1$. This would reduce extra-variability due to incoherence between the code we have developed for lqmHMM and the code developed by \cite{Farcomeni2012} for lqHMM. The simulation scheme and the results we have obtained under Scenario 1 are detailed in the Supplementary Material. 
A second scenario (Scenario 2) has been considered to assess how \textit{lqHMM+LDO} behaves when a non-ignorable missing data process affect the longitudinal responses with respect to the corresponding MAR counterpart, that is the \textit{lqmHMM}. 
Two different intensities for the relationship between the drop-out and the time spent into the study have been considered in order to capture differences between the two model specifications. 

\subsection{Simulation Scenario 2: partially complete data}
Data have been generated from a mixed hidden Markov model with two hidden states ($m = 2$) and three latent drop-out classes ($G = 3$). We have considered longitudinal measures on a sample of $n = 100, 200$ individuals at $T = 5,10$ equally spaced measurement occasions; some individuals drop-out from the study before the planned end, thus presenting incomplete data records. To simulate the time to drop-out, $T_i$, we have considered a discrete distribution with $\Pr(T_i = j) = 1/(T-1) , j = 1, ..., T$, meaning that approximately only $25\%$ ($T=5$) and $11\%$ ($T=10$) of the enrolled individuals do present complete data records. Initial and transition probabilities for the hidden Markov chain have been fixed to 
\begin{align}
\boldsymbol \delta = \left( 0.7, 0.3 \right) \quad  \text{and} \quad  \textbf Q = \left(
\begin{matrix}
0.7 & 0.3 \\
0.3 & 0.7
\end{matrix} \right), 
\end{align}
and two different sets of $\bs \lambda$ parameters have been considered for the missing data process. The former set, $\lambda_{01} = 5, \lambda_{02} = 8.5, \lambda_{1} = -1.1$, has been chosen so that class probabilities are strongly related to the drop-out time (``high informative drop-out scenario''). The latter set, $\lambda_{01} = 1, \lambda_{02} = 2.75, \lambda_{1} = -0.3$, implies that the drop-out time does not strongly influence the LDO class membership 
(``low informative drop-out scenario''). For the longitudinal observations, the following regression model holds for the $h$-th state of the Markov chain and the $g$-th LDO class:
\begin{equation}\label{cap6_simul_model}
Y_{it} = \alpha_{h} + b_g\,  x_{it1} + \beta \, x_{it2} + \varepsilon_{it},
\end{equation}
where $\beta=-0.8$, $x_{it1} \sim \text{N}(1, 3)$ and $\tbf x_{it2} \sim \text{Unif}[0,10]$. State-dependent intercepts have been set to $\alpha_1 = 100$ and $\alpha_2 = 102.5$, while LDO-dependent parameters have been set to $b_1 = 0.5, b_2 = 1.5, b_3 = 3$.
\textit{The difference between state-specific intercepts, that is ($\alpha_1, \alpha_2$), has been set to a lower value than the one considered in Scenario 1, to verify whether aliasing between the $\alpha$s and the $b_g$s may occur in such a scenario.}
\\
Also, we have considered different probability distributions to generate the measurement error, that is a standard Gaussian distribution, a Student $t_{3}$ distribution, and a $\chi_{2}^{2}$ distribution, where the latter two scenarios allow for heavy tailed and skewed data, respectively. 
\\
We have generated $B = 250$ samples and model parameters have been estimated for three quantiles $\tau = (0.25, 0.50, 0.75)$. To evaluate the model performance, the bias and root mean square error (RMSE), over simulations, have been computed for each model parameter. 
\\
As no significant differences have been found between results for $n = 100$ and $n = 200$, Tables \ref{simul_q25}-\ref{simul_q75} show only results obtained under the former scenario for $\tau = 0.25$, $\tau = 0.50$ and $\tau = 0.75$, respectively; results for $n = 200$ are available from the authors upon request. 

{Before going into details, it is worth to contextualize the simulation study and highlight what we expect. As it can be easily observed, the models we are comparing share the same linear predictor and the same overall structure but for the mixture component probabilities. In the \textit{lqHMM+LDO} formulation, these are directly related to the time each individual spent into the study, while, in the \textit{lqmHMM} one, they are constant over individuals. Therefore, we expect that differences between the two approaches will be negligible in the ``low informative drop-out scenario'', that is when setting $\lambda_1 = -0.3$, and more evident in the other one, defined by $\lambda_1 = -1.1$. Indeed, in this latter case, the components of the finite mixture are closely related to $T_i$ and, therefore, \textit{lqHMM+LDO} should be able to recover more accurately individual memberships to LDO classes and ensure more stable estimates for the component-specific locations $b_g$. Obviously, with increasing $T$, the effect of the different mixture components become clearer and, as a result, we should observe that behaviours of the 
\textit{lqmHMM} estimates approach those of \textit{lqHMM+LDO}.}

\begin{center}
Tables \ref{simul_q25}-\ref{simul_q75} about here
\end{center}

{As expected, in all the considered simulation scenarios, the quality of parameter estimates increases as the number of available measures increases both for \textit{lqmHMM} and \textit{lqHMM+LDO}. 
The fixed parameter $\beta$ in the longitudinal data model is always estimated with higher accuracy than the parameters associated to time-constant or time-varying latent variables.
Higher RMSEs are generally observed when moving far from the center of the response distribution, especially for the state-dependent intercepts. \textbf{According to the data generating procedure, these parameters are directly related to the $\tau$-th quantile value of the error distribution and, therefore, in low density regions, such as in the tails, the quality of results generally decreases. With increasing sample sizes and number of measurement occasions the quality of results seems to improve, but for the $\alpha_2$ parameter in the $\chi^2_2$ case for $\tau = 0.75$. Here, both \textit{lqHMM+LDO} and \textit{lqmHMM} algorithms seem to face some difficulties in recovering the true effect, possibly due to some aliasing between time-constant and time-varying random parameters.
However, based on the parameters we have considered for the simulation, the first state of the hidden Markov chain is the most likely one; as a result $\alpha_1$ is generally estimated with some more accuracy than $\alpha_2$.
Lower RMSE are obtained in the case of Gaussian errors when compared to the heavy tailed and the skewed case because of a reduced amount of information, except for the first quartile ($\tau = 0.25$), where $\chi^2$ distributed random errors  ensure more information that lead to a higher quality estimates.}
\\
When comparing results obtained by fitting \textit{lqHMM+LDO} with those coming from the \textit{lqmHMM} formulation, better results are generally obtained, both in terms of bias (more evidently) and variability. As expected, differences are more noticeable in the ``high informative drop-out scenario'' and for those parameters related to the  LDO classes.
Based on the chosen values for $b_g$s, individuals dropping-out earlier in time (i.e. belonging to the first LDO classes) are those with lower values of the longitudinal outcome. Therefore, the missing data process mostly influences the left tail of the response variable distribution (as in the CD4 data example). Therefore, in the left tail, the distinction between LDO classes is more evident and overlapping between components is less likely. 
This can somehow explain the reduced differences between \textit{lqHMM+LDO} and \textit{lqmHMM}, when moving from the left to the right tail of the response distribution. 
Figures \ref{fig_ARI_la1}-\ref{fig_ARI_la5} show the distribution of the adjusted rand indexes over the simulations comparing the true LDO class membership and the estimated allocations obtained through the \textit{lqHMM+LDO} and the \textit{lqmHMM}.
\begin{center}
Figures \ref{fig_ARI_la1}-\ref{fig_ARI_la5} about here
\end{center}
As it is clear, when the missing data process is directly taken into consideration, the estimated LDO class memberships are much more reliable for all the considered simulation scenarios and all the estimated quantiles. This result confirms what we have already noticed when analysing the CD4 data.
Although differences between estimated parameters are negligible, the \textit{lqHMM+LDO} formulation offers more homogeneous groups which, thanks to the estimated $\bs \lambda$ parameters in the LDO class model, are also easier to be interpreted.
}}

\section{Concluding remarks}
In this manuscript we have discussed a class of mixed hidden Markov quantile regression models for longitudinal continuous responses; a general dependence structure is considered by allow the measurement from each statistical units share both time-varying and time constant random parameters, thus providing an extension to the models proposed by \cite{GeraciBottai2007} and \cite{Farcomeni2012}. Both unobserved heterogeneity sources are modelled by using discrete distribution, in a nonparametric fashion, in order to produce a conditional model which should be robust under a series of empirical situations, as shown in the simulation study. Since unobserved heterogeneity in this context may arise due to omitted covariates or be influenced by patterns of drop-out, we allow the time-constant random parameters to have a distribution which is dependent on the observed pattern of drop-out (ie the number of measurements for each individual) through a drop-out related ordered latent class, as suggested by \cite{Roy2003} and \cite{Roy2008}. The simulation study and the re-analysis of a well known benchmark dataset, the CD4 cells data of \cite{Kaslow1987}, give quite encouraging results, showing how the proposed models can be easily applied to complex data structures.

\bibliographystyle{plainnat}
\bibliography{biblio}

\pagebreak

\section*{Tables}

\begin{table}[!ht]
\centering
\caption{Simulation study for $\tau = 0.25$. Bias and RMSE for longitudinal parameter estimates under the \textit{lqmHMM} and \textit{lqHMM+LDO} formulation.} 
\scalebox{0.90}{

\begin{tabular}{llrrrrrrrrrrrrrrrrrrrrrrrrr}
 \toprule
  	
  &	& \multicolumn{2}{c}{$N$} &
  	\multicolumn{2}{c}{$t_3$} &
  	\multicolumn{2}{c}{$\chi_2^2$} \\
  	
&& \multicolumn{1}{c}{\textit{lqmHMM}} &  \multicolumn{1}{c}{\textit{lqHMM+LDO}}	& \multicolumn{1}{c}{\textit{lqmHMM}} & \multicolumn{1}{c}{\textit{lqHMM+LDO}}& \multicolumn{1}{c}{\textit{lqmHMM}} &\multicolumn{1}{c}{\textit{lqHMM+LDO}}\\
\hline
\vspace{-0.5cm}
\multirow{14}{*}{\rotatebox{90}{$\bs \lambda = (1,2.75,-0.3)$}}\\

&&\multicolumn{7}{l}{$\mathtt{T = 5}$}\\

&		$\alpha_1$	&	-0.001	(0.25)	&	0.026	(0.27)	&	0.014	(0.39)	&	0.059	(0.35)	&	-0.042	(0.20)	&	-0.032	(0.20)	\\
&		$\alpha_2$	&	-0.606	(0.83)	&	-0.628	(0.85)	&	-0.998	(1.19)	&	-0.984	(1.19)	&	-0.090	(0.24)	&	-0.087	(0.25)	\\
&		$\beta$	&	0.003	(0.04)	&	-0.008	(0.04)	&	-0.001	(0.05)	&	-0.002	(0.05)	&	0.000	(0.03)	&	0.001	(0.03)	\\
&		$b_1$	&	0.000	(0.05)	&	-0.002	(0.05)	&	-0.003	(0.06)	&	-0.006	(0.06)	&	-0.005	(0.03)	&	-0.006	(0.03)	\\
&		$b_2$	&	-0.001	(0.05)	&	-0.004	(0.05)	&	0.006	(0.06)	&	0.005	(0.06)	&	-0.007	(0.04)	&	-0.008	(0.04)	\\
&		$b_3$	&	-0.004	(0.06)	&	-0.005	(0.06)	&	0.002	(0.08)	&	-0.001	(0.07)	&	0.005	(0.05)	&	0.002	(0.05)	\\

	\cline{2-10}							
					
													&	&\multicolumn{7}{l}{$\mathtt{T = 10}$}	\\

&		$\alpha_1$	&	0.099	(0.25)	&	-0.071	(0.23)	&	0.188	(0.41)	&	-0.236	(0.59)	&	-0.105	(0.43)	&	-0.094	(0.16)	\\
&		$\alpha_2$	&	0.463	(0.68)	&	-0.427	(0.69)	&	0.884	(1.13)	&	-0.867	(1.12)	&	-0.153	(0.38)	&	-0.142	(0.22)	\\
&		$\beta$	&	-0.001	(0.03)	&	-0.001	(0.03)	&	-0.004	(0.04)	&	0.001	(0.04)	&	0.002	(0.02)	&	0.001	(0.02)	\\
&		$b_1$	&	-0.010	(0.05)	&	0.007	(0.05)	&	-0.006	(0.06)	&	0.004	(0.06)	&	0.006	(0.06)	&	0.005	(0.03)	\\
&		$b_2$	&	-0.015	(0.04)	&	0.002	(0.10)	&	-0.003	(0.10)	&	0.004	(0.08)	&	0.004	(0.10)	&	0.004	(0.02)	\\
&		$b_3$	&	-0.019	(0.04)	&	0.008	(0.10)	&	-0.003	(0.09)	&	-0.002	(0.09)	&	0.005	(0.13)	&	0.003	(0.02)	\\

\hline
\hline
\vspace{-0.5cm}

\multirow{15}{*}{\rotatebox{90}{$\bs \lambda = (5,8.5,-1.1)$}}\\

&&\multicolumn{7}{l}{$\mathtt{T = 5}$}\\	

&		$\alpha_1$	&	-0.106	(0.33)	&	-0.075	(0.32)	&	-0.137	(0.46)	&	-0.099	(0.47)	&	-0.108	(0.21)	&	-0.093	(0.21)	\\
&		$\alpha_2$	&	-0.913	(1.08)	&	-0.897	(1.07)	&	-1.164	(1.32)	&	-1.143	(1.30)	&	-0.185	(0.37)	&	-0.195	(0.40)	\\
&		$\beta$	&	0.002	(0.04)	&	0.001	(0.04)	&	0.005	(0.05)	&	0.002	(0.05)	&	0.002	(0.03)	&	0	(0.03)	\\
&		$b_1$	&	-0.007	(0.08)	&	0	(0.05)	&	-0.020	(0.11)	&	-0.004	(0.05)	&	0.002	(0.06)	&	0.006	(0.03)	\\
&		$b_2$	&	-0.051	(0.22)	&	-0.02	(0.15)	&	-0.055	(0.24)	&	-0.001	(0.10)	&	-0.007	(0.12)	&	0.006	(0.04)	\\
&		$b_3$	&	-0.272	(0.63)	&	-0.26	(0.59)	&	-0.275	(0.63)	&	-0.238	(0.57)	&	-0.177	(0.44)	&	-0.166	(0.42)	\\

	\cline{2-10}							
					
													&&\multicolumn{7}{l}{$\mathtt{T = 10}$}	\\				
													
&		$\alpha_1$	&	-0.076	(0.23)	&	-0.035	(0.22)	&	-0.124	(0.34)	&	-0.071	(0.31)	&	-0.095	(0.15)	&	-0.078	(0.14)	\\
&		$\alpha_2$	&	-0.407	(0.62)	&	-0.375	(0.59)	&	-0.824	(1.04)	&	-0.756	(0.99)	&	-0.145	(0.23)	&	-0.121	(0.18)	\\
&		$\beta$	&	0.000	(0.03)	&	-0.002	(0.03)	&	0.002	(0.04)	&	0.001	(0.04)	&	-0.001	(0.02)	&	-0.001	(0.02)	\\
&		$b_1$	&	0.011	(0.05)	&	0.008	(0.05)	&	-0.004	(0.07)	&	-0.006	(0.07)	&	0.004	(0.03)	&	0.003	(0.03)	\\
&		$b_2$	&	0.015	(0.04)	&	0.013	(0.04)	&	0.009	(0.05)	&	0.007	(0.05)	&	0.003	(0.02)	&	0.002	(0.02)	\\
&		$b_3$	&	0.017	(0.04)	&	0.013	(0.04)	&	0.007	(0.04)	&	0.004	(0.04)	&	0.004	(0.02)	&	0.001	(0.02)	\\

 \bottomrule
\bottomrule
\end{tabular}\label{simul_q25}

}

\end{table}

\begin{table}[!ht]
\centering
\caption{Simulation study for $\tau = 0.50$. Bias and RMSE for longitudinal parameter estimates under the \textit{lqmHMM} and \textit{lqHMM+LDO} formulation.} 
\scalebox{0.90}{

\begin{tabular}{llrrrrrrrrrrrrrrrrrrrrrrrrr}
 \toprule
  	
  &	& \multicolumn{2}{c}{$N$} &
  	\multicolumn{2}{c}{$t_3$} &
  	\multicolumn{2}{c}{$\chi_2^2$} \\
  	
&& \multicolumn{1}{c}{\textit{lqmHMM}} &  \multicolumn{1}{c}{\textit{lqHMM+LDO}}	& \multicolumn{1}{c}{\textit{lqmHMM}} & \multicolumn{1}{c}{\textit{lqHMM+LDO}}& \multicolumn{1}{c}{\textit{lqmHMM}} &\multicolumn{1}{c}{\textit{lqHMM+LDO}}\\
\hline
\vspace{-0.5cm}
\multirow{14}{*}{\rotatebox{90}{$\bs \lambda = (1,2.75,-0.3)$}}\\

&&\multicolumn{7}{l}{$\mathtt{T = 5}$}\\	

&		$\alpha_1$	&	0.011	(0.25)	&	-0.076	(0.24)	&	0.017	(0.28)	&	-0.060	(0.27)	&	-0.218	(0.50)	&	-0.260	(0.46)	\\
&		$\alpha_2$	&	-0.022	(0.74)	&	-0.255	(0.40)	&	-0.092	(0.48)	&	-0.271	(0.46)	&	-0.130	(0.72)	&	-0.322	(0.46)	\\
&		$\beta$	&	0.001	(0.04)	&	-0.002	(0.04)	&	0.001	(0.04)	&	0.000	(0.04)	&	0.002	(0.05)	&	0.003	(0.05)	\\
&		$b_1$	&	0.009	(0.05)	&	0.016	(0.05)	&	0.007	(0.06)	&	0.007	(0.05)	&	-0.002	(0.08)	&	-0.016	(0.05)	\\
&		$b_2$	&	-0.078	(0.27)	&	-0.003	(0.13)	&	-0.071	(0.25)	&	-0.004	(0.11)	&	-0.058	(0.19)	&	-0.028	(0.11)	\\
&		$b_3$	&	-0.180	(0.48)	&	-0.012	(0.19)	&	-0.148	(0.45)	&	0.000	(0.10)	&	-0.097	(0.33)	&	-0.021	(0.15)	\\

	\cline{2-10}							
					
													&	&\multicolumn{7}{l}{$\mathtt{T = 10}$}	\\

&		$\alpha_1$	&	0.016	(0.23)	&	0.060	(0.18)	&	0.023	(0.30)	&	0.077	(0.21)	&	-0.248	(0.46)	&	-0.107	(0.58)	\\
&		$\alpha_2$	&	-0.116	(0.24)	&	-0.085	(0.23)	&	-0.097	(0.25)	&	-0.058	(0.23)	&	-0.339	(0.44)	&	-0.327	(0.49)	\\
&		$\beta$	&	0.001	(0.03)	&	0.000	(0.03)	&	0.000	(0.03)	&	0.000	(0.03)	&	-0.002	(0.03)	&	0.003	(0.03)	\\
&		$b_1$	&	0.010	(0.06)	&	-0.002	(0.04)	&	0.009	(0.09)	&	-0.004	(0.04)	&	0.014	(0.14)	&	-0.013	(0.05)	\\
&		$b_2$	&	0.003	(0.04)	&	-0.003	(0.03)	&	0.009	(0.10)	&	-0.003	(0.04)	&	0.010	(0.16)	&	-0.015	(0.05)	\\
&		$b_3$	&	0.003	(0.04)	&	-0.002	(0.03)	&	0.001	(0.03)	&	-0.005	(0.03)	&	-0.010	(0.04)	&	-0.002	(0.04)	\\

\hline
\hline
\vspace{-0.5cm}

\multirow{15}{*}{\rotatebox{90}{$\bs \lambda = (5,8.5,-1.1)$}}\\

&&\multicolumn{7}{l}{$\mathtt{T = 5}$}\\	

&		$\alpha_1$	&	0.041	(0.25)	&	0.066	(0.26)	&	0.056	(0.26)	&	0.081	(0.27)	&	-0.071	(0.55)	&	0.603	(0.63)	\\
&		$\alpha_2$	&	-0.148	(0.42)	&	-0.131	(0.35)	&	-0.096	(0.40)	&	-0.080	(0.40)	&	-0.159	(0.67)	&	0.468	(0.52)	\\
&		$\beta$	&	0.000	(0.04)	&	0.000	(0.04)	&	0.001	(0.04)	&	0.001	(0.04)	&	-0.004	(0.05)	&	0.002	(0.03)	\\
&		$b_1$	&	0.004	(0.05)	&	0.000	(0.05)	&	0.002	(0.05)	&	-0.002	(0.05)	&	-0.012	(0.07)	&	0.003	(0.05)	\\
&		$b_2$	&	-0.042	(0.17)	&	-0.008	(0.06)	&	-0.013	(0.07)	&	-0.009	(0.06)	&	-0.018	(0.09)	&	0.006	(0.04)	\\
&		$b_3$	&	-0.298	(0.67)	&	-0.282	(0.64)	&	-0.191	(0.51)	&	-0.193	(0.50)	&	-0.096	(0.47)	&	0.008	(0.03)	\\

	\cline{2-10}							
					
													&&\multicolumn{7}{l}{$\mathtt{T = 10}$}	\\

&		$\alpha_1$	&	-0.118	(0.41)	&	-0.047	(0.21)	&	-0.064	(0.29)	&	-0.011	(0.19)	&	-0.395	(0.52)	&	-0.340	(0.50)	\\
&		$\alpha_2$	&	-0.212	(0.46)	&	-0.206	(0.33)	&	-0.185	(0.32)	&	-0.146	(0.26)	&	-0.487	(0.55)	&	-0.475	(0.53)	\\
&		$\beta$	&	-0.002	(0.02)	&	-0.003	(0.02)	&	0.003	(0.03)	&	0.002	(0.03)	&	0.001	(0.03)	&	0.000	(0.03)	\\
&		$b_1$	&	0.012	(0.05)	&	0.011	(0.05)	&	0.004	(0.05)	&	0.003	(0.05)	&	0.001	(0.05)	&	0.005	(0.05)	\\
&		$b_2$	&	0.011	(0.07)	&	0.014	(0.04)	&	0.007	(0.04)	&	0.006	(0.04)	&	0.005	(0.04)	&	0.007	(0.04)	\\
&		$b_3$	&	0.014	(0.09)	&	0.017	(0.04)	&	0.011	(0.03)	&	0.008	(0.03)	&	0.009	(0.04)	&	0.007	(0.04)	\\

 \bottomrule
\bottomrule
\end{tabular}\label{simul_q50}

}

\end{table}

\begin{table}[!ht]
\centering
\caption{Simulation study for $\tau = 0.75$. Bias and RMSE for longitudinal parameter estimates under the \textit{lqmHMM} and \textit{lqHMM+LDO} formulation.} 
\scalebox{0.90}{

\begin{tabular}{llrrrrrrrrrrrrrrrrrrrrrrrrr}
 \toprule
  \toprule
  	
  &	& \multicolumn{2}{c}{$N$} &
  	\multicolumn{2}{c}{$t_3$} &
  	\multicolumn{2}{c}{$\chi_2^2$} \\
  	
&& \multicolumn{1}{c}{\textit{lqmHMM}} &  \multicolumn{1}{c}{\textit{lqHMM+LDO}}	& \multicolumn{1}{c}{\textit{lqmHMM}} & \multicolumn{1}{c}{\textit{lqHMM+LDO}}& \multicolumn{1}{c}{\textit{lqmHMM}} &\multicolumn{1}{c}{\textit{lqHMM+LDO}}\\

\hline\hline
\vspace{-0.5cm}

\multirow{14}{*}{\rotatebox{90}{$\bs \lambda = (1,2.75,-0.3)$}}\\

&&\multicolumn{7}{l}{$\mathtt{T = 5}$}\\

&	$\alpha_1$	&	-0.001	(0.25)	&	0.026	(0.27)	&	0.014	(0.39)	&	0.059	(0.35)	&	-0.042	(0.20)	&	-0.032	(0.20)	\\
&	$\alpha_2$	&	-0.606	(0.83)	&	-0.628	(0.85)	&	-0.998	(1.19)	&	-0.984	(1.19)	&	-0.090	(0.24)	&	-0.087	(0.25)	\\
&	$\beta$	&	0.003	(0.04)	&	-0.008	(0.04)	&	-0.001	(0.05)	&	-0.002	(0.05)	&	0.000	(0.03)	&	0.001	(0.03)	\\
&	$b_1$	&	0.000	(0.05)	&	-0.002	(0.05)	&	-0.003	(0.06)	&	-0.006	(0.06)	&	-0.005	(0.03)	&	-0.006	(0.03)	\\
&	$b_2$	&	-0.001	(0.05)	&	-0.004	(0.05)	&	0.006	(0.06)	&	0.005	(0.06)	&	-0.007	(0.04)	&	-0.008	(0.04)	\\
&	$b_3$	&	-0.004	(0.06)	&	-0.005	(0.06)	&	0.002	(0.08)	&	-0.001	(0.07)	&	0.005	(0.05)	&	0.002	(0.05)	\\

	\cline{2-10}

													&	&\multicolumn{7}{l}{$\mathtt{T = 10}$}	\\

&	$\alpha_1$	&	-0.099	(0.25)	&	-0.100	(0.16)	&	0.188	(0.41)	&	-0.236	(0.59)	&	-0.105	(0.43)	&	-0.094	(0.16)	\\
&	$\alpha_2$	&	-0.463	(0.68)	&	-0.149	(0.21)	&	0.884	(1.13)	&	-0.867	(1.12)	&	-0.153	(0.38)	&	-0.142	(0.22)	\\
&	$\beta$	&	0.001	(0.03)	&	0.000	(0.02)	&	-0.004	(0.04)	&	0.001	(0.04)	&	0.002	(0.02)	&	0.001	(0.02)	\\
&	$b_1$	&	0.010	(0.05)	&	0.005	(0.03)	&	-0.006	(0.06)	&	0.004	(0.06)	&	0.006	(0.06)	&	0.005	(0.03)	\\
&	$b_2$	&	0.015	(0.04)	&	0.003	(0.03)	&	-0.003	(0.10)	&	0.004	(0.08)	&	0.004	(0.10)	&	0.004	(0.02)	\\
&	$b_3$	&	0.019	(0.04)	&	0.007	(0.02)	&	-0.003	(0.09)	&	-0.002	(0.09)	&	0.005	(0.13)	&	0.003	(0.02)	\\

\hline\hline
\vspace{-0.5cm}
\multirow{15}{*}{\rotatebox{90}{$\bs \lambda = (5,8.5,-1.1)$}}\\

&&\multicolumn{7}{l}{$\mathtt{T = 5}$}\\	

&	$\alpha_1$	&	-0.106	(0.33)	&	-0.075	(0.32)	&	-0.137	(0.46)	&	-0.099	(0.47)	&	-0.108	(0.21)	&	-0.093	(0.21)	\\
&	$\alpha_2$	&	-0.913	(1.08)	&	-0.897	(1.07)	&	-1.164	(1.32)	&	-1.143	(1.30)	&	-0.185	(0.37)	&	-0.195	(0.40)	\\
&	$\beta$	&	0.002	(0.04)	&	0.001	(0.04)	&	0.005	(0.05)	&	0.002	(0.05)	&	0.002	(0.03)	&	0.000	(0.03)	\\
&	$b_1$	&	-0.007	(0.08)	&	0.000	(0.05)	&	-0.020	(0.11)	&	-0.004	(0.05)	&	0.002	(0.06)	&	0.006	(0.03)	\\
&	$b_2$	&	-0.051	(0.22)	&	-0.020	(0.15)	&	-0.055	(0.24)	&	-0.001	(0.10)	&	-0.007	(0.12)	&	0.006	(0.04)	\\
&	$b_3$	&	-0.272	(0.63)	&	-0.260	(0.59)	&	-0.275	(0.63)	&	-0.238	(0.57)	&	-0.177	(0.44)	&	-0.166	(0.42)	\\

	\cline{2-10}							
					
													&&\multicolumn{7}{l}{$\mathtt{T = 10}$}	\\			
													
&	$\alpha_1$	&	-0.076	(0.23)	&	-0.035	(0.22)	&	-0.124	(0.34)	&	-0.071	(0.31)	&	-0.095	(0.15)	&	-0.078	(0.14)	\\
&	$\alpha_2$	&	-0.407	(0.62)	&	-0.375	(0.59)	&	-0.824	(1.04)	&	-0.756	(0.99)	&	-0.145	(0.23)	&	-0.121	(0.18)	\\
&	$\beta$	&	0.000	(0.03)	&	-0.002	(0.03)	&	0.002	(0.04)	&	0.001	(0.04)	&	-0.001	(0.02)	&	-0.001	(0.02)	\\
&	$b_1$	&	0.011	(0.05)	&	0.008	(0.05)	&	-0.004	(0.07)	&	-0.006	(0.07)	&	0.004	(0.03)	&	0.003	(0.03)	\\
&	$b_2$	&	0.015	(0.04)	&	0.013	(0.04)	&	0.009	(0.05)	&	0.007	(0.05)	&	0.003	(0.02)	&	0.002	(0.02)	\\
&	$b_3$	&	0.017	(0.04)	&	0.013	(0.04)	&	0.007	(0.04)	&	0.004	(0.04)	&	0.004	(0.02)	&	0.001	(0.02)	\\

 \bottomrule
\bottomrule
\end{tabular}\label{simul_q75}

}

\end{table}

\begin{table}[t]
\caption{CD4 data. BIC values for lqmHMM for different choices of $m$ and $G$ at different quantiles.}
\centering
\begin{tabular}{lrrrrrrrr}
\toprule
 \toprule
&\multicolumn{6}{c}{Mixture Components}\\
Hidden States  & \multicolumn{1}{c}{1} & \multicolumn{1}{c}{2} & \multicolumn{1}{c}{3} & \multicolumn{1}{c}{4} &\multicolumn{1}{c}{5} & \multicolumn{1}{c}{6}\\ 
\midrule
\multicolumn{2}{l}{$\mathtt{\tau = 0.25}$}\\
1 & 3940.50 & 3530.48 & 3408.42 & 3312.98 & 3286.67 & 3298.81 \\ 
  2 & 3292.25 & 2923.33 & 2780.49 & 2772.64 & 2760.39 & 2772.33 \\ 
  3 & 2963.50 & 2747.21 & 2676.26 & 2637.39 & 2619.05 & 2627.48 \\ 
  4 & 2757.55 & 2660.63 & 2541.63 & 2510.71 & \textbf{2478.88} & 2487.22 \\ 
  5 & 2688.98 & 2527.39 & 2504.99 & 2494.63 & 2486.83 & 2507.90 \\

 \multicolumn{2}{l}{$\mathtt{\tau = 0.50}$}\\
 
 1 & 3434.26 & 3014.58 & 2898.11 & 2850.34 & 2833.41 & 2874.04 \\ 
  2 & 2733.04 & 2522.56 & 2412.12 & 2381.26 & 2374.18 & 2381.82 \\ 
  3 & 2523.15 & 2345.76 & 2280.27 & 2266.98 & 2256.28 & 2268.15 \\ 
  4 & 2410.07 & 2268.98 & 2236.89 & \textbf{2233.57 }& 2236.29 & 2236.84 \\ 
  5 & 2377.80 & 2298.48 & 2272.29 & 2266.93 & 2267.97 & 2278.48 \\ 
 
\multicolumn{2}{l}{$\mathtt{\tau = 0.75}$}\\ 
1 & 3491.69 & 3134.31 & 2986.87 & 2951.28 & 2951.44 & 2940.33 \\ 
  2 & 2823.44 & 2549.81 & 2495.15 & 2455.24 & 2441.21 & 2444.51 \\ 
  3 & 2470.11 & 2337.64 & 2290.17 & 2242.39 & 2244.89 & 2256.24 \\ 
  4 & 2370.11 & 2307.19 & \textbf{2228.76} & 2231.06 & 2240.10 & 2230.01 \\ 
  5 & 2356.16 & 2295.80 & 2251.37 & 2252.42 & 2252.06 & 2274.55 \\ 
\bottomrule
\bottomrule
\end{tabular}
\label{cap5_tab_CD4_BIC}

\end{table}

\begin{table}[t]
\caption{CD4 data. Estimated longitudinal model parameters for lqmHMM at different quantiles. $95\%$ bootstrap confidence intervals are reported within brackets.}
\centering

\begin{small}

\begin{tabular}{lrrrrrrrr}
\toprule
 \toprule
& \multicolumn{2}{c}{$\mathtt{\tau = 0.25}$} &  \multicolumn{2}{c}{$\mathtt{\tau = 0.50}$} &  \multicolumn{2}{c}{$\mathtt{\tau = 0.75}$}\\
& \multicolumn{2}{c}{$[m=4, G=5]$} &  \multicolumn{2}{c}{$[m=4, G=4]$} &  \multicolumn{2}{c}{$[m=4, G=3]$}\\
\midrule
$\alpha_1$	&	$5.593$	&	$(5.403	;	5.677)$	&	$6.054$	&	$(5.994	;	6.133)$	&	$6.203$	&	$(6.071	;	6.273)$	\\
$\alpha_2$	&	$6.124$	&	$(6.066	;	6.166)$	&	$6.432$	&	$(6.368	;	6.530)$	&	$6.580$	&	$(6.517	;	6.628)$	\\
$\alpha_3$	&	$6.540$	&	$(6.489	;	6.587)$	&	$6.750$	&	$(6.689	;	6.837)$	&	$6.876$	&	$(6.804	;	6.934)$	\\
$\alpha_4$	&	$6.915$	&	$(6.847	;	6.995)$	&	$7.055$	&	$(7.023	;	7.231)$	&	$7.256$	&	$(7.168	;	7.373)$	\\
Age	&	$0.000$	&	$(-0.004	;	0.002)$	&	$0.004$	&	$(-0.001	;	0.008)$	&	$0.000$	&	$(-0.005	;	0.005)$	\\
Drugs	&	$0.044$	&	$(0.000	;	0.092)$	&	$0.057$	&	$(-0.014	;	0.110)$	&	$0.061$	&	$(0.003	;	0.113)$	\\
Packs	&	$0.056$	&	$(0.041	;	0.071)$	&	$0.043$	&	$(0.015	;	0.054)$	&	$0.044$	&	$(0.015	;	0.062)$	\\
Partners	&	$0.006$	&	$(0.001	;	0.012)$	&	$0.005$	&	$(0.001	;	0.012)$	&	$0.011$	&	$(0.003	;	0.016)$	\\
CES-D	&	$-0.004$	&	$(-0.005	;	-0.001)$	&	$-0.004$	&	$(-0.006	;	-0.002)$	&	$-0.004$	&	$(-0.006	;	-0.002)$	\\
$\text{Time}_{\text{sero}}$	&	$-0.175$	&	$(-0.206	;	-0.150)$	&	$-0.140$	&	$(-0.164	;	-0.114)$	&	$-0.123$	&	$(-0.145	;	-0.102)$	\\
$\sigma_b$	&	$0.219$	&	$(0.200	;	0.360)$	&	$0.134$	&	$(0.105	;	0.165)$	&	$0.102$	&	$(0.088	;	0.133)$	\\

\bottomrule
\bottomrule
\end{tabular}

\end{small}
\label{cap5_tab_CD4_estimLong}
\end{table}

\begin{sidewaystable}[ht]
\caption{CD4 data. Estimated initial and transition probabilities for lqmHMM, at different quantiles. $95\%$ bootstrap confidence intervals are reported within brackets.  }
\centering

\begin{tabular}{x{0.7cm}rrrrrrrrr}
\toprule
 \toprule
 & \multicolumn{2}{c}{1} & \multicolumn{2}{c}{2} & \multicolumn{2}{c}{3} & \multicolumn{2}{c}{4} \\ 

\midrule

\multicolumn{2}{l}{$\mathtt{\tau = 0.25}$}\\
\hdashline
$\delta$	&	$0.083$	&	$(0.036	;	0.134)$	&	$0.408$	&	$(0.316	;	0.512)$	&	$0.411$	&	$(0.304	;	0.501)$	&	$0.098$	&	$(0.055	;	0.157)$	\\
\hdashline
1	&	$0.284$	&	$(0.116	;	0.466)$	&	$0.700$	&	$(0.510	;	0.860)$	&	$0.000$	&	$(0.000	;	0.025)$	&	$0.016$	&	$(0.000	;	0.069)$	\\
2	&	$0.083$	&	$(0.034	;	0.125)$	&	$0.675$	&	$(0.574	;	0.754)$	&	$0.232$	&	$(0.153	;	0.334)$	&	$0.010$	&	$(0.000	;	0.032)$	\\
3	&	$0.031$	&	$(0.008	;	0.062)$	&	$0.126$	&	$(0.077	;	0.179)$	&	$0.787$	&	$(0.705	;	0.844)$	&	$0.055$	&	$(0.019	;	0.119)$	\\
4	&	$0.011$	&	$(0.000	;	0.033)$	&	$0.044$	&	$(0.000	;	0.090)$	&	$0.015$	&	$(0.000	;	0.095)$	&	$0.930$	&	$(0.854	;	0.977)$	\\

\multicolumn{2}{l}{$\mathtt{\tau = 0.50}$}\\
\hdashline
$\delta$	&	0.120	&	$(0.059	;	0.242)$	&	$0.434$	&	$(0.306	;	0.543)$	&	$0.326$	&	$(0.213	;	0.449)$	&	$0.120$	&	$(0.059	;	0.162)$	\\
\hdashline
1	&	$0.927$	&	$(0.775	;	1.000)$	&	$0.073$	&	$(0.000	;	0.216)$	&	$0.000$	&	$(0.000	;	0.034)$	&	$0.000$	&	$(0.000	;	0.016)$	\\
2	&	$0.107$	&	$(0.052	;	0.174)$	&	$0.830$	&	$(0.734	;	0.925)$	&	$0.063$	&	$(0.000	;	0.144)$	&	$0.000$	&	$(0.000	;	0.016)$	\\
3	&	$0.013$	&	$(0.000	;	0.062)$	&	$0.055$	&	$(0.000	;	0.108)$	&	$0.898$	&	$(0.838	;	0.961)$	&	$0.034$	&	$(0.000	;	0.071)$	\\
4	&	$0.043$	&	$(0.000	;	0.079)$	&	$0.000$	&	$(0.000	;	0.041)$	&	$0.019$	&	$(0.000	;	0.092)$	&	$0.938$	&	$(0.866	;	0.994)$	\\

\multicolumn{2}{l}{$\mathtt{\tau = 0.75}$}\\
\hdashline
$\delta$	&	$0.119$	&	$(0.041	;	0.192)$	&	$0.359$	&	$(0.224	;	0.497)$	&	$0.367$	&	$(0.232	;	0.488)$	&	$0.155$	&	$(0.098	;	0.217)$	\\
\hdashline
1	&	$0.861$	&	$(0.757	;	0.958)$	&	$0.139$	&	$(0.041	;	0.243)$	&	$0.000$	&	$(0.000	;	0.000)$	&	$0.000$	&	$(0.000	;	0.000)$	\\
2	&	$0.125$	&	$(0.069	;	0.194)$	&	$0.810$	&	$(0.701	;	0.885)$	&	$0.065$	&	$(0.000	;	0.171)$	&	$0.000$	&	$(0.000	;	0.022)$	\\
3	&	$0.021$	&	$(0.000	;	0.062)$	&	$0.094$	&	$(0.031	;	0.184)$	&	$0.858$	&	$(0.770	;	0.921)$	&	$0.026$	&	$(0.000	;	0.063)$	\\
4	&	$0.019$	&	$(0.000	;	0.048)$	&	$0.000$	&	$(0.000	;	0.044)$	&	$0.088$	&	$(0.007	;	0.192)$	&	$0.894$	&	$(0.782	;	0.965)$	\\

\bottomrule
\bottomrule
\end{tabular}

\label{cap5_tab_CD4_estimQ}

\end{sidewaystable}

\begin{small}
\begin{table}[htb]
\centering
\caption{CD4 data. Number of individuals in the study at each time occasion.}
\begin{tabular}{cccccccccccccccccc}
\toprule
\toprule
Visit & 1 & 2 & 3 & 4 & 5 & 6 & 7 & 8 & 9 & 10 & 11 & 12 \\ 
\midrule
& 369 & 364 & 340 & 315 & 268 & 225 & 173 & 133 &  92 &  54 &  33 &  10 \\ 
\bottomrule
\bottomrule
\end{tabular}\label{cap6_tab_missing}
\end{table}
\end{small}

\begin{small}
\begin{table}[h]
\caption{CD4 data. BIC values for lqHMM+LDO for different choices of $m$ and $G$ at different quantiles.}
\centering
\begin{tabular}{lrrrrrrr}
\toprule
\toprule
&\multicolumn{4}{c}{LDO classes}\\
Hidden States  & \multicolumn{1}{c}{1} & \multicolumn{1}{c}{2} & \multicolumn{1}{c}{3} & \multicolumn{1}{c}{4} &\multicolumn{1}{c}{5} \\ 

\midrule
\multicolumn{2}{l}{$\mathtt{\tau = 0.25}$}\\

1	&	3940.50	&	3525.57	&	3400.96	&	3305.67	&	3279.57	\\
2	&	3292.25	&	2919.41	&	2773.66	&	2761.84	&	2755.28	\\
3	&	2963.50	&	2741.63	&	2669.37	&	2636.84	&	2612.29	\\
4	&	2757.55	&	2660.38	&	2537.89	&	2509.15	&	2522.63	\\
5	&	2688.98	&	2551.40	&	2522.53	&	2474.39	&	\textbf{2460.52}	\\

 \multicolumn{2}{l}{$\mathtt{\tau = 0.50}$}\\
 
1	&	3434.26	&	3010.15	&	2895.63	&	2847.35	&	2829.65	\\
2	&	2733.04	&	2517.26	&	2406.67	&	2377.40	&	2369.86	\\
3	&	2523.15	&	2343.65	&	2280.14	&	2266.03	&	2259.80	\\
4	&	2410.07	&	2265.06	&	2233.48	&	\textbf{2231.14}	&	2233.14	\\
5	&	2377.80	&	2291.68	&	2265.61	&	2259.65	&	2244.27	\\

\multicolumn{2}{l}{$\mathtt{\tau = 0.75}$}\\ 

1	&	3491.69	&	3137.21	&	2987.71	&	2953.09	&	2953.71	\\
2	&	2823.44	&	2551.39	&	2491.32	&	2453.73	&	2448.10	\\
3	&	2470.11	&	2335.57	&	2287.38	&	2240.26	&	2242.14	\\
4	&	2370.11	&	2308.85	&	\textbf{2225.93}	&	\textbf{2203.98}	&	2223.70	\\
5	&	2356.17	&	2290.69	&	2248.35	&	2240.17	&	2242.98	\\

\bottomrule
\bottomrule
\end{tabular}

\label{cap6_tab_CD4_BIC}

\end{table}
\end{small}

\begin{table}[!ht]
\caption{CD4 data. Estimated longitudinal model parameters for lqHMM+LDO at different quantiles. $95\%$ bootstrap confidence intervals are reported within brackets.}
\centering

\begin{small}

\begin{tabular}{lrrrrrrrr}
\toprule
 \toprule
& \multicolumn{2}{c}{$\mathtt{\tau = 0.25}$} &  \multicolumn{2}{c}{$\mathtt{\tau = 0.50}$} &  \multicolumn{2}{c}{$\mathtt{\tau = 0.75}$}\\
& \multicolumn{2}{c}{$[m = 5, G=5]$} & \multicolumn{2}{c}{$[m = 4, G=4]$} &\multicolumn{2}{c}{$[m = 4, G=3]$} \\
\midrule

$\alpha_1$	&	$5.046$	&	$(3.937	;	5.286)$	&	$6.043$	&	$(5.931	;	6.114)$	&	$6.198$	&	$(6.069	;	6.282)$	\\
$\alpha_2$	&	$5.880$	&	$(5.730	;	5.918)$	&	$6.416$	&	$(6.323	;	6.502)$	&	$6.579$	&	$(6.512	;	6.628)$	\\
$\alpha_3$	&	$6.193$	&	$(6.126	;	6.256)$	&	$6.719$	&	$(6.647	;	6.825)$	&	$6.872$	&	$(6.801	;	6.934)$	\\
$\alpha_4$	&	$6.582$	&	$(6.508	;	6.634)$	&	$7.040$	&	$(6.973	;	7.215)$	&	$7.243$	&	$(7.167	;	7.370)$	\\
$\alpha_5$	&	$6.936$	&	$(6.846	;	7.026)$	&		&				&		&			
\\
Age	&	$-0.004$	&	$(-0.007	;	0.000)$	&	$0.004$	&	$(-0.001	;	0.007)$	&	$0.000$	&	$(-0.004	;	0.005)$	\\
Drugs	&	$0.048$	&	$(-0.013	;	0.124)$	&	$0.072$	&	$(-0.006	;	0.145)$	&	$0.064$	&	$(0.007	;	0.115)$	\\
Packs	&	$0.032$	&	$(0.024	;	0.051)$	&	$0.042$	&	$(0.014	;	0.054)$	&	$0.044$	&	$(0.018	;	0.064)$	\\
Partners	&	$0.011$	&	$(0.005	;	0.016)$	&	$0.005$	&	$(0.000	;	0.012)$	&	$0.011$	&	$(0.002	;	0.016)$	\\
CES-D	&	$-0.003$	&	$(-0.006	;	-0.001)$	&	$-0.004$	&	$(-0.006	;	-0.002)$	&	$-0.004$	&	$(-0.006	;	-0.002)$	\\
$\text{Time}_{\text{sero}}$	&	$-0.157$	&	$(-0.187	;	-0.127)$	&	$-0.146$	&	$(-0.175	;	-0.119)$	&	$-0.131$	&	$(-0.155	;	-0.108)$	\\

\bottomrule
\bottomrule
\end{tabular}

\end{small}
\label{cap6_tab_long}
\end{table}

\begin{sidewaystable}[!h]
\caption{CD4 data. Estimated initial and transition probabilities for lqHMM+LDO, at different quantiles. $95\%$ bootstrap confidence intervals are reported within brackets.}
\centering

\scalebox{0.95}{
\begin{tabular}{x{0.7cm}rrrrrrrrrrrr}
\toprule
 \toprule
 & \multicolumn{2}{c}{1} & \multicolumn{2}{c}{2} & \multicolumn{2}{c}{3} & \multicolumn{2}{c}{4} &  \multicolumn{2}{c}{5} \\ 

\midrule

\multicolumn{2}{l}{$\mathtt{\tau = 0.25}$}\\
\hdashline
$\delta$	&	$0.010$	&	$(0.000	;	0.026)$	&	$0.125$	&	$(0.054	;	0.187)$	&	$0.365$	&	$(0.254	;	0.477)$	&	$0.358$	&	$(0.263	;	0.470)$	&	$0.142$	&	$(0.073	;	0.212)$	\\
\hdashline
1	&	$0.264$	&	$(0.000	;	0.610)$	&	$0.588$	&	$(0.000	;	0.855)$	&	$0.092$	&	$(0.000	;	0.616)$	&	$0.000$	&	$(0.000	;	0.427)$	&	$0.056$	&	$(0.000	;	0.285)$	\\
2	&	$0.049$	&	$(0.000	;	0.115)$	&	$0.471$	&	$(0.167	;	0.791)$	&	$0.480$	&	$(0.143	;	0.797)$	&	$0.000$	&	$(0.000	;	0.000)$	&	$0.000$	&	$(0.000	;	0.000)$	\\
3	&	$0.027$	&	$(0.000	;	0.050)$	&	$0.139$	&	$(0.052	;	0.212)$	&	$0.592$	&	$(0.430	;	0.706)$	&	$0.225$	&	$(0.130	;	0.376)$	&	$0.017$	&	$(0.000	;	0.049)$	\\
4	&	$0.006$	&	$(0.000	;	0.022)$	&	$0.030$	&	$(0.000	;	0.077)$	&	$0.165$	&	$(0.103	;	0.229)$	&	$0.737$	&	$(0.640	;	0.815)$	&	$0.062$	&	$(0.013	;	0.128)$	\\
5	&	$0.004$	&	$(0.000	;	0.015)$	&	$0.008$	&	$(0.000	;	0.030)$	&	$0.047$	&	$(0.000	;	0.096)$	&	$0.095$	&	$(0.000	;	0.180)$	&	$0.845$	&	$(0.764	;	0.947)$	\\

\multicolumn{2}{l}{$\mathtt{\tau = 0.50}$}\\
\cdashline{1-9}
$\delta$	&	$0.118$	&	$(0.048	;	0.215)$	&	$0.411$	&	$(0.295	;	0.544)$	&	$0.341$	&	$(0.221	;	0.454)$	&	$0.130$	&	$(0.066	;	0.180)$	&		&				\\
\cdashline{1-9}
1	&	$0.939$	&	$(0.753	;	1.000)$	&	$0.061$	&	$(0.000	;	0.246)$	&	$0.000$	&	$(0.000	;	0.000)$	&	$0.000$	&	$(0.000	;	0.010)$	&		&				\\
2	&	$0.104$	&	$(0.060	;	0.193)$	&	$0.846$	&	$(0.711	;	0.912)$	&	$0.050$	&	$(0.000	;	0.155)$	&	$0.000$	&	$(0.000	;	0.020)$	&		&				\\
3	&	$0.015$	&	$(0.000	;	0.053)$	&	$0.049$	&	$(0.004	;	0.101)$	&	$0.901$	&	$(0.836	;	0.955)$	&	$0.036$	&	$(0.000	;	0.084)$	&		&				\\
4	&	$0.041$	&	$(0.000	;	0.082)$	&	$0.000$	&	$(0.000	;	0.036)$	&	$0.045$	&	$(0.000	;	0.126)$	&	$0.914$	&	$(0.835	;	0.981)$	&		&				\\

\multicolumn{2}{l}{$\mathtt{\tau = 0.75}$}\\
\cdashline{1-9}
$\delta$	&	$0.122$	&	$(0.040	;	0.205)$	&	$0.361$	&	$(0.219	;	0.510)$	&	$0.360$	&	$(0.221	;	0.488)$	&	$0.157$	&	$(0.094	;	0.225)$	&		&				\\
\cdashline{1-9}
1	&	$0.866$	&	$(0.749	;	0.973)$	&	$0.134$	&	$(0.027	;	0.251)$	&	$0.000$	&	$(0.000	;	0.000)$	&	$0.000$	&	$(0.000	;	0.000)$	&		&				\\
2	&	$0.122$	&	$(0.065	;	0.199)$	&	$0.813$	&	$(0.695	;	0.894)$	&	$0.065$	&	$(0.000	;	0.168)$	&	$0.000$	&	$(0.000	;	0.023)$	&		&				\\
3	&	$0.020$	&	$(0.000	;	0.055)$	&	$0.099$	&	$(0.039	;	0.194)$	&	$0.854$	&	$(0.758	;	0.916)$	&	$0.027$	&	$(0.000	;	0.064)$	&		&				\\
4	&	$0.018$	&	$(0.000	;	0.045)$	&	$0.000$	&	$(0.000	;	0.045)$	&	$0.091$	&	$(0.012	;	0.207)$	&	$0.891$	&	$(0.765	;	0.960)$	&		&				\\

\bottomrule
\bottomrule
\end{tabular}
}

\label{cap6_tab_HMM}

\end{sidewaystable}

\begin{table}[!h]
\caption{CD4 data. Estimated LDO-dependent parameters in the longitudinal data model for lqHMM+LDO at different quantiles. $95\%$ bootstrap confidence intervals are reported within brackets.}
\centering
\begin{small}

\begin{tabular}{lrrrrrrrrrr}
\toprule
\toprule
& \multicolumn{2}{c}{lqHMM+LDO} & & \multicolumn{2}{c}{lqmHMM}\\
\midrule

\multicolumn{2}{l}{$\mathtt{\tau = 0.25}$}\\ 

$b_1$	&	$-0.849$	&	$(-1.568	;	-0.802)$	&	&	$-0.740$	&	$(-1.079	;	-0.662)$	\\
$b_2$	&	$-0.434$	&	$(-0.447	;	-0.401)$	&	&	$-0.300$	&	$(-0.324	;	-0.256)$	\\
$b_3$	&	$-0.220$	&	$(-0.245	;	-0.203)$	&	&	$-0.164$	&	$(-0.181	;	-0.133)$	\\
$b_4$	&	$-0.123$	&	$(-0.141	;	-0.099)$	&	&	$-0.053$	&	$(-0.095	;	-0.035)$	\\
$b_5$	&	$-0.020$	&	$(-0.041	;	-0.004)$	&	&	$0.026$	&	$(-0.011	;	0.045)$	\\

\multicolumn{2}{l}{$\mathtt{\tau = 0.50}$}\\

$b_1$	&	$-0.502$	&	$(-0.617	;	-0.370)$	&	&	$-0.497$	&	$(-0.667	;	-0.452)$	\\
$b_2$	&	$-0.175$	&	$(-0.204	;	-0.158)$	&	&	$-0.176$	&	$(-0.200	;	-0.155)$	\\
$b_3$	&	$-0.071$	&	$(-0.104	;	-0.061)$	&	&	$-0.070$	&	$(-0.098	;	-0.056)$	\\
$b_4$	&	$0.026$	&	$(-0.027	;	0.037)$	&	&	$0.033$	&	$(-0.023	;	0.047)$	\\

\multicolumn{2}{l}{$\mathtt{\tau = 0.75}$}\\ 

$b_1$	&	$-0.328$	&	$(-0.423	;	-0.297)$	&	&	$-0.327$	&	$(-0.414	;	-0.287)$	\\
$b_2$	&	$-0.114$	&	$(-0.130	;	-0.093)$	&	&	$-0.113$	&	$(-0.131	;	-0.093)$	\\
$b_3$	&	$-0.001$	&	$(-0.020	;	0.020)$	&	&	$0.003$	& $(-0.023	;	0.019)$	\\

\bottomrule
\bottomrule
\end{tabular}

\end{small}
\label{cap6_tab_ldoLongPar}
\end{table}

\begin{table}[!hb]
\caption{CD4 data. Estimated LDO class model parameters for lqHMM+LDO at different quantiles. $95\%$ bootstrap confidence intervals are reported within brackets.}
\centering
\begin{small}

\begin{tabular}{lrrrrrrrrrr}
\toprule
\toprule
& \multicolumn{2}{c}{$\mathtt{\tau = 0.25}$} & \multicolumn{2}{c}{$\mathtt{\tau = 0.50}$} &\multicolumn{2}{c}{$\mathtt{\tau = 0.75}$} \\
& \multicolumn{2}{c}{$[m = 5, G=5]$} & \multicolumn{2}{c}{$[m = 4, G=4]$} &\multicolumn{2}{c}{$[m = 4, G=3]$} \\
\midrule

$\lambda_{01}$	&	$-2.385$	&	$(-3.583	;	-1.383)$	&	$-1.062$	&	$(-2.112	;	-0.241)$	&	$-0.374$	&	$(-1.388	;	0.615)$	\\
$\lambda_{02}$	&	$-0.082$	& $(-1.159	;	0.993)$	&	$1.113$	&	$(0.013	;	2.102)$	&	$2.739$	&	$(1.295	;	4.379)$	\\
$\lambda_{03}$	&	$1.555$	&	$(0.514	;	2.627)$	&	$4.089$	&	$(2.002	;	5.299)$	&		&				\\
$\lambda_{04}$	&	$3.116$	&	$(1.926	;	4.388)$	&		&				&		&				\\
$\lambda_{1}$	&	$-0.174$	&	$(-0.290	;	-0.059)$	&	$-0.193$	&	$(-0.318	;	-0.065)$	&	$-0.184$	&	$(-0.324	;	-0.066)$	\\

\bottomrule
\bottomrule
\end{tabular}
\end{small}
\\
\label{cap6_tab_ldoModel}

\end{table}

\clearpage

\section*{Figures}
%
%
%
%
%
%
%
%

\begin{figure}[ht]
\caption{CD4 data. Response variable distribution at each time occasion.}
\centering
\includegraphics[scale=0.45]{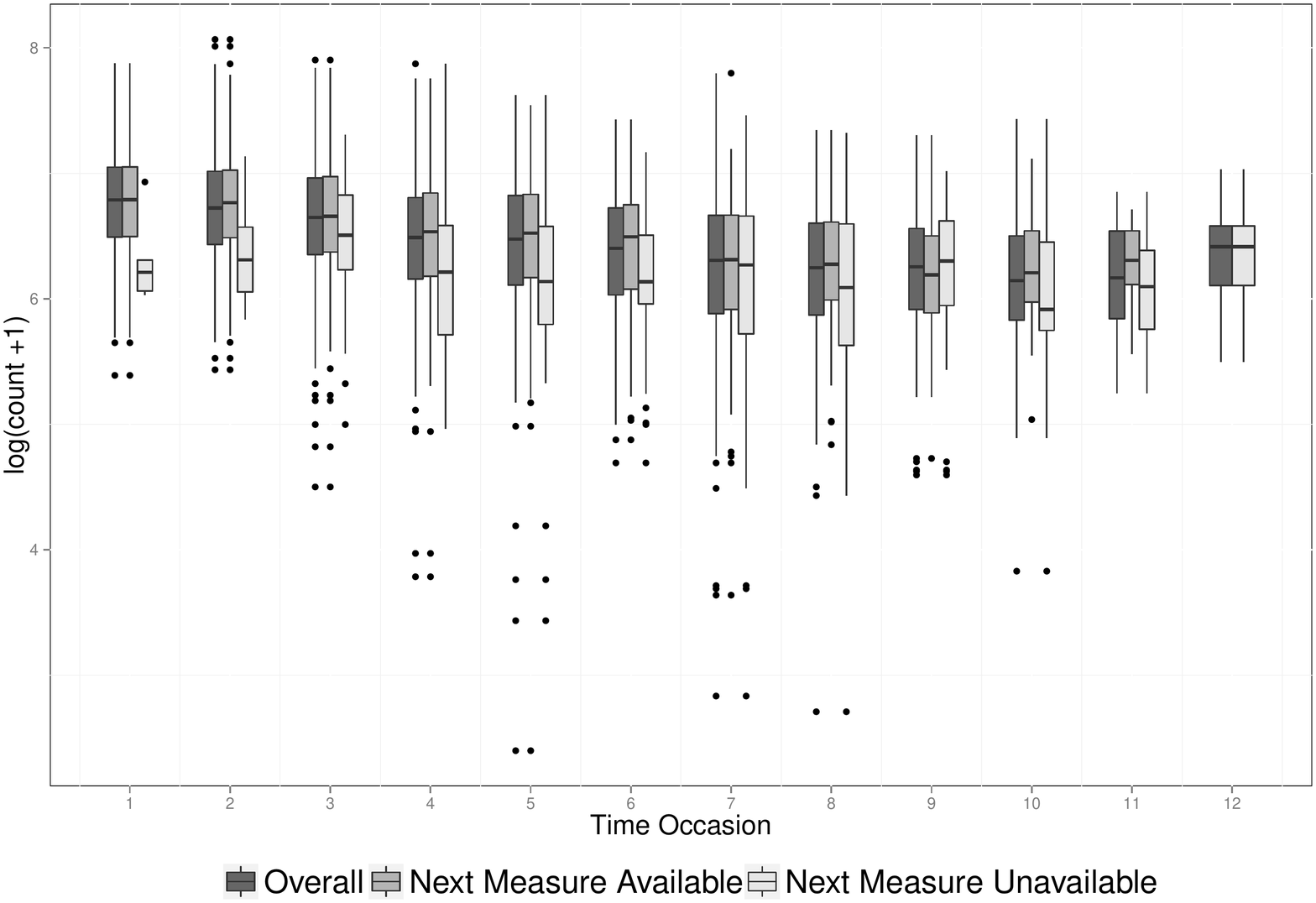} 

\label{cap6_fig_respDist_vs_TDO}
\end{figure}

%
%
%
\begin{figure}[!b]
\caption{CD4 data. Longitudinal trajectories within LDO classes for \textit{lqHMM+LDO} at different quantiles}
\centering
\includegraphics[scale = 0.70]{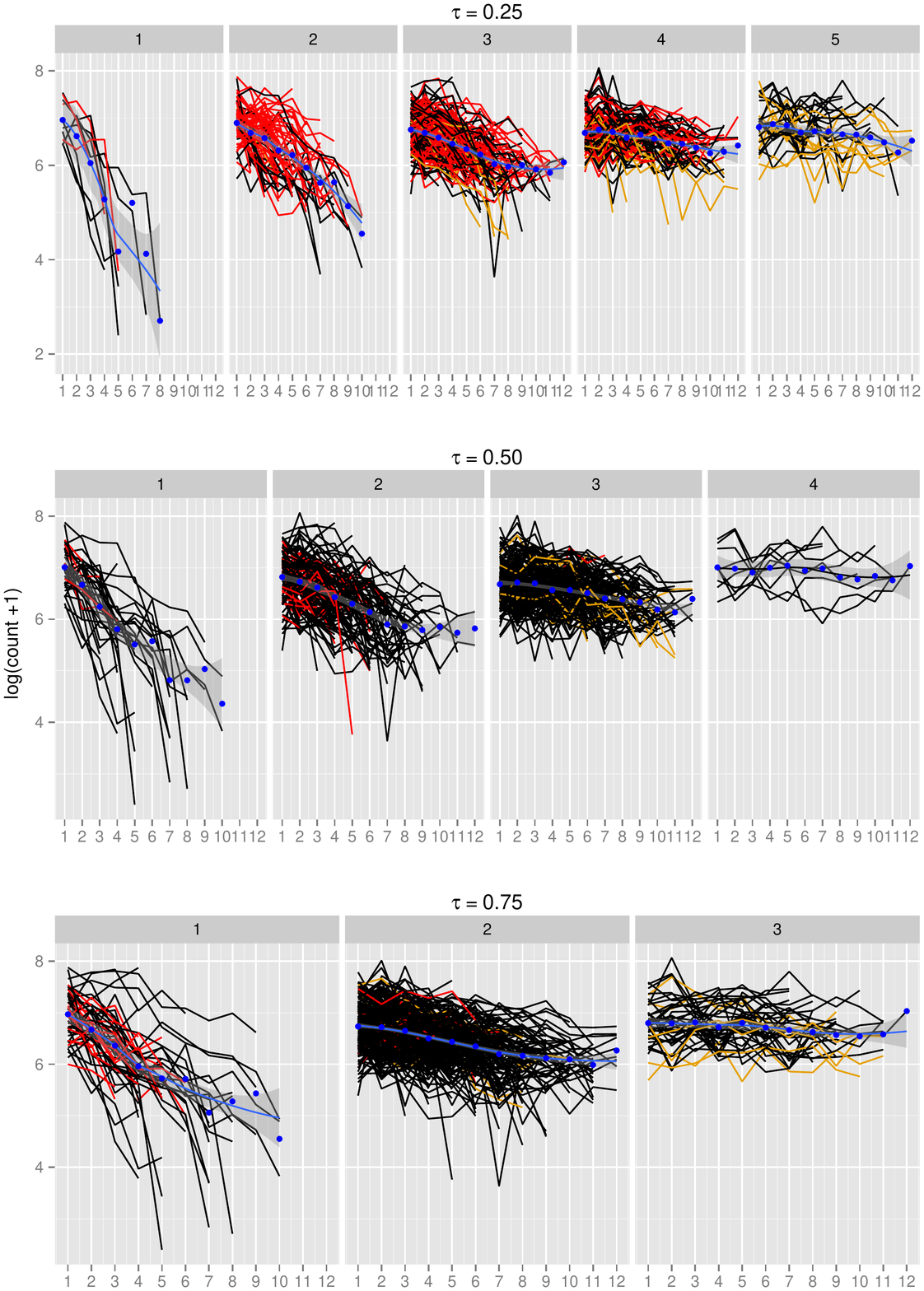}
\label{fig_spaghetti_unique}
\end{figure}

\begin{figure}[!b]
\caption{Simulation study. $\bs \lambda = (1,2.75,-0.3)$. Distribution of the adjusted rand index for \textit{lqHMM+LDO} (light grey) and \textit{lqmHMM} (dark grey) in the different simulation scenarios at different quantiles}
\centering
\includegraphics[scale = 0.25]{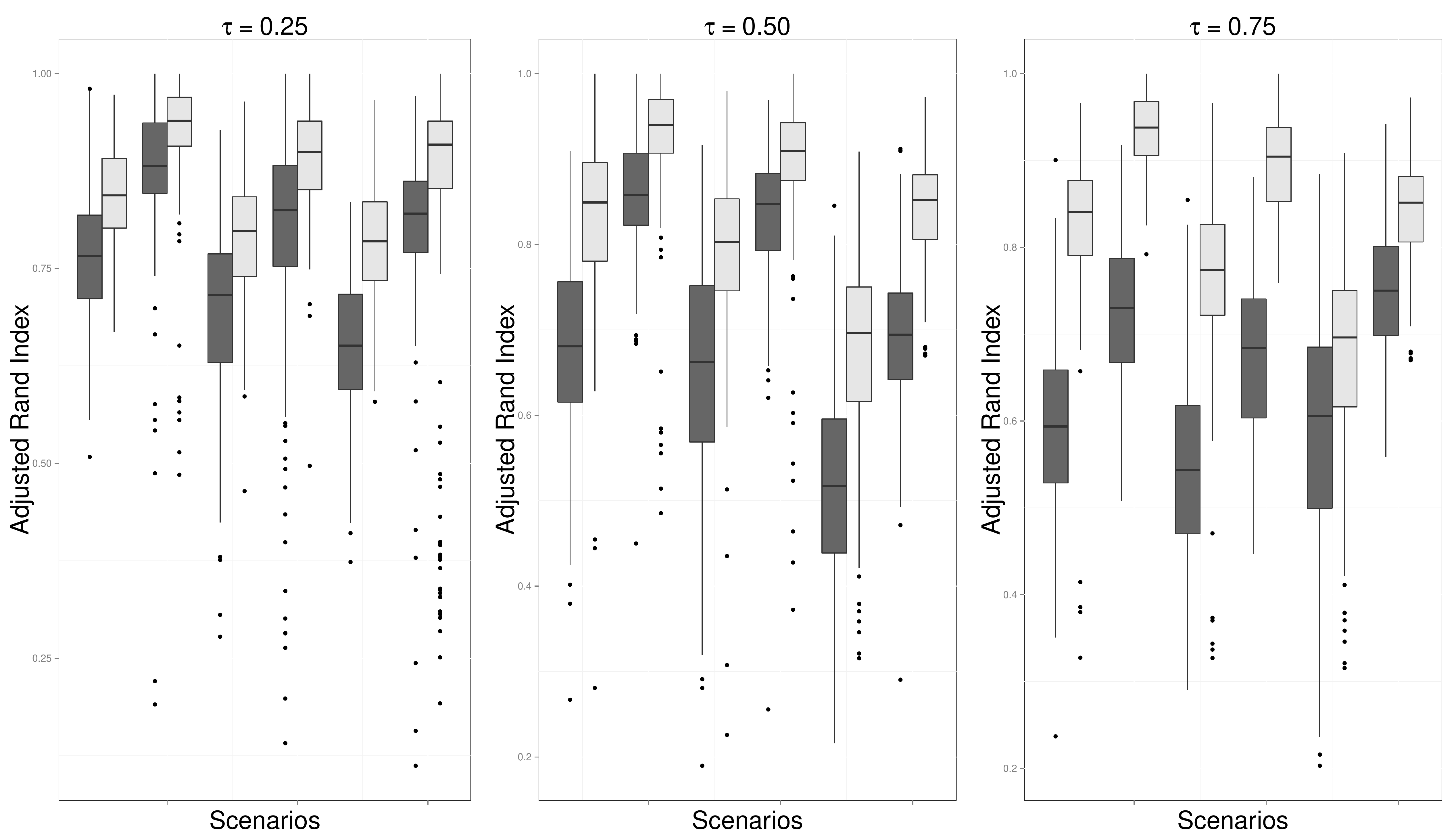}
\label{fig_ARI_la1}
\end{figure}

\begin{figure}[!b]
\caption{Simulation study. $\bs \lambda = (5,8.5,-1.1)$. Distribution of the adjusted rand index for \textit{lqHMM+LDO} (light grey) and \textit{lqmHMM} (dark grey) in the different simulation scenarios at different quantiles}
\centering
\includegraphics[scale = 0.25]{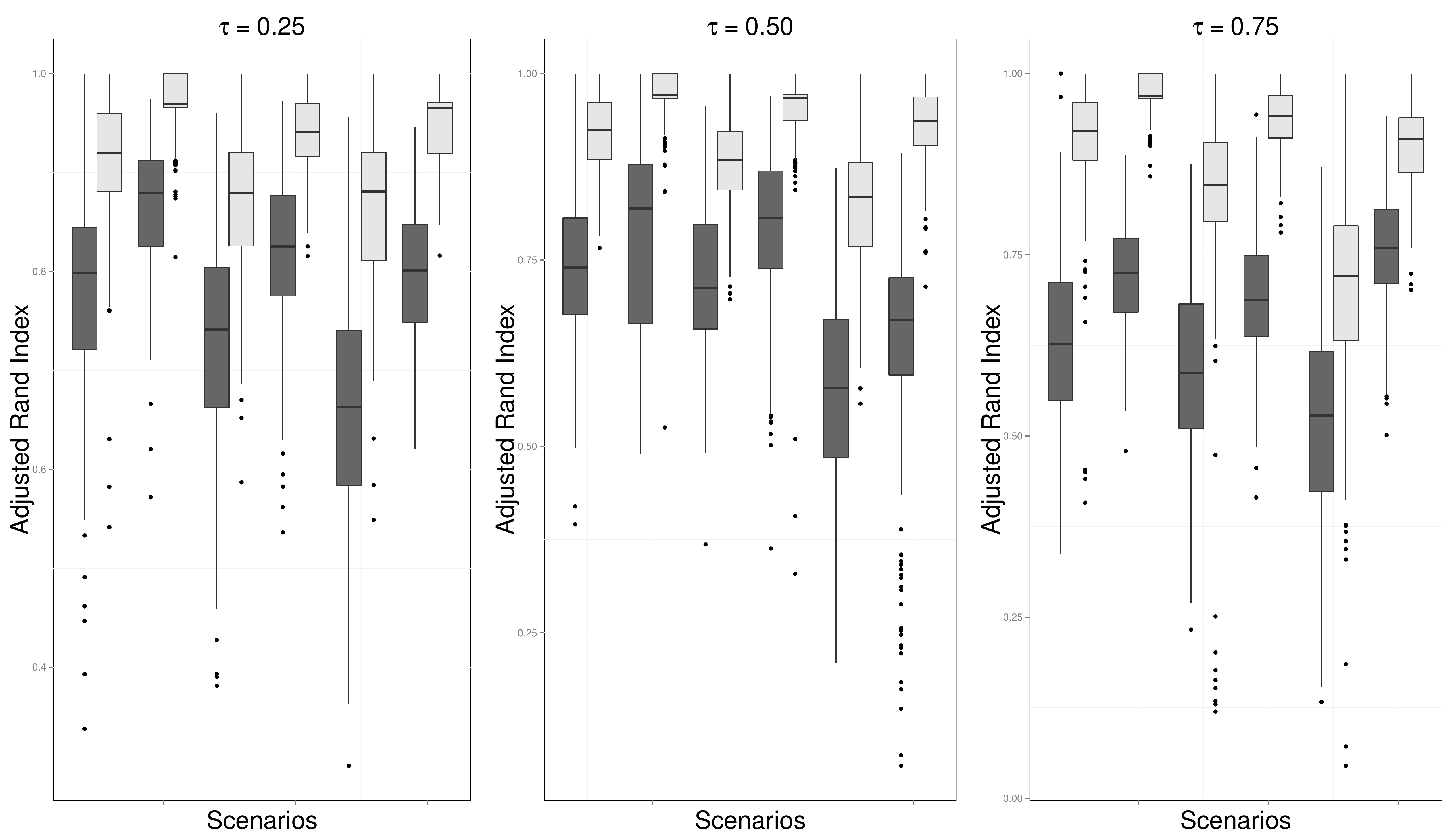}
\label{fig_ARI_la5}
\end{figure}

\end{document}


\maketitle
\onehalfspacing
\section{ML estimation}
Parameter estimates for the \textit{lqmHMM} and the \textit{lqHMM+LDO} are obtained by using the Baum-Welch algorithm \citep{McDonaldZucchini1997}.
As before, we suppress the $\tau$ indexing of the model parameters. We will refer to LDO classes with the generic term 'components'', to align the terminology of \textit{lqmHMM} and \textit{lqHMM+LDO}, and use $\pi_{ig}$ in place of $\pi_{ig} (T_i \mid \bs \lambda, \tau)$ for the \textit{lqHMM+LDO} formulation, while $\pi_{ig} = \pi_{g}, \forall i = 1, \dots, n$ for the \textit{lqmHMM}.
Let $u_{i}(h) = \mathbb I \left[S_{it} = h \right]$ denote the indicator variable for the $i$-th individual in the $h$-th state at time occasion $t$ and let $u_{it}(k,h) = \mathbb I \left[ S_{it-1} = k, S_{it} = h \right]$ indicate whether individual $i$ moves from the $k$-th state at time occasion $t-1$ to the $h$-th one at time occasion $t$. 
Moreover, $\zeta_{ig}$ is the indicator variable for the $i$-th unit belonging to the $g$-th component. Starting from the definition of the complete data log-likelihood,

\begin{align} \label{auxiliary_NPML}
 \ell_c(\bs \Phi \mid \tbf y, \tbf T, \tbf S, \bs \zeta,\tau) &  \propto
\sum _{i = 1} ^ n \bigg\{
\sum_{h = 1}^m u_{i1}(h) \log \delta_{h} +
\sum_{t=2}^{T_i} \sum_{h,k=1}^m u_{it}(k,h) \log q_{kh} + 
\sum_{g = 1}^G \zeta_{ig} \log \pi_{ig} +
\nonumber \\[0.1cm]
& - T_i \log (\sigma) - 
\sum _{t = 1}^{T_i}  \sum_{h=1}^m  
\sum_{g = 1}^G u_{it}(h) \zeta_{ig}
\rho_{\tau} \left[ 
		\frac{ y_{it} - \mu_{it}[S_{it} = h, \tbf b_g] }{\sigma} 
\right]  
 \bigg\},
\end{align}
parameter estimates are derived by using an EM algorithm. 
In the E-step we take the expected value of the complete data log-likelihood \eqref{auxiliary_NPML}, given the observed data and the current parameter estimates; we refer to such quantity as $Q(\bs \Phi \mid \bs \Phi^{(r)})$. This amounts to replacing the indicator variables by the corresponding (posterior) expected values given by the following expressions
\begin{align*}
& \hat u_{it}(h \mid \tau) = {\rm E}_{{\bs \Phi}^{(r)}}\left[u_{it}(h)\mid \tbf y_{i}\right] \\
 & \hat u_{it}(k,h \mid \tau) = {\rm E}_{{\bs \Phi}^{(r)}}\left[u_{it}(k,h )\mid \tbf y_{i}\right] \\
& \hat{\zeta}_{ig}(\tau) = {\rm E}_{{\bs \Phi}^{(r)}}\left[\zeta_{ig} \mid {\bf y}_{i}\right] \\
 & \hat u_{it}(h \mid g, \tau) = {\rm E}_{{\bs \Phi}^{(r)}}\left[u_{it}(h )\mid \zeta_{ig}=1, \tbf y_{i}\right] \\
\end{align*}
The calculation of these terms is greatly simplified by considering forward and backward variables; see \cite{Baum1970}. In the present framework, forward variables, $a_{it} (h, g \mid \tau)$, are defined as the joint density of the longitudinal measures up to time $t$, for a generic individual ending up in the $h$-th state, conditional on the $g$-th mixture component:
\begin{align}\label{cap5_fwdRecursion}
a_{it}(h,g \mid \tau) =
f \left[y_{i1:t}, S_{it} = h \mid g, \tau \right ].
\end{align}
Similarly, the backward variables $b_{it} (h, g \mid \tau)$ represent the probability of the longitudinal sequence from occasion $t+1$ to the last observation $T_i$, conditional on being in the $h$-th state at time $t$ and the $g$-th mixture component:
\begin{align}\label{cap5_bwdRecursion}
b_{it}(h,g \mid \tau) =
f \big[ y_{it+1:T_i} \mid S_{it} = h, g, \tau \big].
\end{align}
Both terms can be computed recursively as
\begin{align*}
& a_{i1}(h,g \mid \tau)  = \delta_h f_{y \mid sb} \big[ y_{i1} \mid S_{i1} = h, g, \tau \big]  \\
&a_{it}(h,g \mid \tau)  =
\sum_{k=1}^m a_{it-1}(k,g \mid \tau) q_{kh} f_{y \mid sb} \big[ y_{it} \mid S_{it} = h, g, \tau \big] \\
& b_{iT_{i}}(h,g \mid \tau) = 1 \\
& b_{it-1}(h,g \mid \tau) = \sum _{k=1}^m
b_{it}(k,\textbf b_g) q_{hk} f_{y \mid sb} \big[ y_{it} \mid S_{it} = k, g, \tau \big].
\end{align*}
Once these variables have been defined, the E-step can be performed straightforwardly. Considering the indicator variables for both the hidden Markov process and the finite mixture as missing data, conditional expected values, given the observed data and the current  parameter estimates, can be easily computed as 
\begin{align}
& \hat u_{it}(h \mid \tau) = \frac
{
\sum_g
a_{it}(h,g \mid \tau)
b_{it}(h,g \mid \tau)
\pi_g
}
{
\sum_{h} \sum_g
a_{it}(h,g \mid \tau)
b_{it}(h,g \mid \tau)
\pi_g
}, 		\nonumber \\[0.4cm]
 & \hat u_{it}(k,h \mid \tau) =
\frac{
\sum_g
a_{it-1}(k,g \mid \tau)
q_{kh} \,
f_{y \mid sb}\left(y_{it} \mid S_{it} = h,_g, \tau \right)
b_{it}(h,g \mid \tau)
\pi_g
} {
\sum_{h k} \sum_g 
a_{it-1}(k,g \mid \tau)
q_{kh} \,
f_{y \mid sb}\left(y_{it} \mid S_{it} = h,g, \tau \right)
b_{it}(h,g \mid \tau)
\pi_g
}, 	 \nonumber \\[0.4cm]
& \hat{\zeta}_{ig}(\tau) = \frac{\sum_{h = 1}^m a_{iT_{i}}(h,g \mid \tau) \pi_g}{\sum_{l = 1}^G \sum_{h = 1}^m a_{iT_{i}}(h,l \mid \tau) \pi_l}	\nonumber \\[0.4cm]
& u_{it}(h \mid g, \tau) = \frac{a_{it}(h,g \mid \tau) b_{it}(h,g \mid \tau) \pi_g}{\sum_{k = 1}^m a_{it}(k,g \mid \tau) b_{it}(k,g \mid \tau) \pi_g}.
\end{align}
The first three rows correspond to the posterior expected values of state and component membership indicators described so far, while the last one represents the conditional posterior probability of being in state $h$ at time occasion $t$, given the individual belongs to the $g$-th component of the finite mixture.

In the M-step, model parameter estimates are derived by maximizing $Q(\bs \Phi \mid \bs \Phi^{(r)})$ with respect to $\bs \Phi$. Given the separability of the parameter spaces for the longitudinal and the missing data process, the maximization can be partitioned into independent sub-problems, thus considerably simplifying the computation. 
Standard results for basic HMMs are valid for the initial and the transition probability estimates,
\begin{align}
\hat \delta _h = \frac{\sum _{i = 1}^n \hat u_{i1}(h \mid \tau)}{n},
\quad
\hat q_{kh} = \frac {\sum _{i = 1}^n \sum_{t=1}^{T_i} \hat u_{it}(k,h \mid \tau)}{\sum _{i = 1}^n \sum_{t=1}^{T_i} \sum _{h = 1}^m \hat u_{it}(h,k \mid \tau)}, \quad h, k = 1, \dots, m.
\end{align}
\\
Longitudinal model parameters, say $\bs \psi$, are estimated by solving a weighted estimating equation, with weights given by the posterior probabilities of the hidden Markov process and the finite mixture. More specifically, the updates are obtained as
\begin{align}
\hat{\bs \psi}=\arg \min_{\bs \psi} 
\sum_{i=1}^n \sum _{t = 1}^{T_i}  \sum_{h=1}^m \sum_{g = 1}^G
\hat \zeta_{ig}(\tau) \hat u_{it}(h \mid g, \tau) 
\rho_{\tau} \left[ \frac{y_{it} - \mu_{it} [S_{it} = h, \tbf b_g]
}
{\sigma} \right].
\label{eq_longEstim_NPML}
\end{align}
By extending the block algorithm proposed by \cite{Farcomeni2012} for the \textit{lqHMM}, three different steps have to be alternated. Noticing that $S_{it} = h$ implies $\bs\alpha_{s_{it}} = \bs\alpha_h$, fixed parameters $\bs \beta$ are estimated by solving 
\begin{equation}
\hat{\bs \beta}=\arg \min_{\bs \beta} \sum_{i=1}^n \sum _{t = 1}^{T_i}  \sum_{h=1}^m \sum_{g = 1}^G \hat \zeta_{ig}(\tau) \hat u_{it}(h \mid g, \tau) 
\rho_{\tau} \left[ \tilde y_{it} - \tbf x_{it} \bs \beta \right],
\end{equation}
where $\tilde y_{it} = [y_{it} - \tbf z_{it} ^\prime \hat{\tbf b}_g - \tbf w_{it} ^\prime 	\hat{\bs \alpha}_h]$. 
\\
State-dependent parameters ${\bs \alpha}_h, h = 1, ...,m$ are updated via
\begin{equation}
\hat{\bs \alpha_h}=\arg \min_{\bs \alpha_h} \sum_{i=1}^n \sum _{t = 1}^{T_i}  \sum_{h=1}^m \sum_{g = 1}^G \hat \zeta_{ig}(\tau) \hat u_{it}(h \mid g, \tau) 
\rho_{\tau} \left[ \tilde y_{it} -  \bs \alpha_h \right],
\end{equation}
where $\tilde y_{it} = [y_{it} - \tbf x_{it} ^\prime \hat{\bs \beta} - \tbf z_{it} ^\prime \hat{\tbf b}_g]$.
\\
The locations $\tbf b_g$ are computed by solving
\begin{equation}
\hat{\tbf b}_g=\arg \min_{\tbf b_g} \sum_{i=1}^n \sum _{t = 1}^{T_i}  \sum_{h=1}^m \sum_{g = 1}^G \hat \zeta_{ig}(\tau) \hat u_{it}(h \mid g, \tau) 
\rho_{\tau} \left[ \tilde y_{it} -  \tbf z_{it} \tbf b_g \right], 
\end{equation}
with $\tilde y_{it} = [y_{it} - \tbf x_{it} ^\prime \hat{\bs \beta} - \tbf w_{it} ^\prime \hat{\bs \alpha}_h]$.
\\
With regards to the component probabilities, closed form expressions are available for 
the \textit{lqmHMM} formulation: 
\begin{equation}
\hat \pi_g = \frac{1}{n} \sum_{i = 1}^n \hat \zeta_{ig}(\tau).
\label{eq_pgEstim}
\end{equation}
For the \textit{lqHMM+LDO} specification, parameters in the LDO class model are estimated via a constrained optimization of the weighed likelihood for the cumulative logit model, that is
\begin{align}
\hat{\bs \lambda}=& \arg \max_{\bs \lambda} \sum_{g=1}^G \hat{\zeta}_{ig}(\tau)
\left\{
\left[\frac{\exp(\lambda_0 + \lambda_g T_i)}{1+ \exp(\lambda_0 + \lambda_g T_i)}
\right] - 
\left[\frac{\exp(\lambda_0 + \lambda_{g-1} T_i)}{1+ \exp(\lambda_0 + \lambda_{g-1} T_i)}
\right] \right\}.
\end{align}
Finally, the scale parameter of the AL distribution is estimated by
\begin{equation}
\hat \sigma = \frac{1}{\sum_{i=1}^n T_i} \sum_{i = 1}^n \sum_{t=1}^{T_i} \sum_{h=1}^m \sum_{g=1}^G \hat \zeta_i(g\mid \tau) \hat{u}_{it}(h\mid g, \tau) \rho_\tau
\left(
y_{it} - \mu_{it}(S_{it} = h, \tbf b_g)
\right).
\end{equation}
The E- and the M-step of the algorithm are iterated until convergence, that is until the (relative) difference between subsequent likelihood values is lower than an arbitrary small amount $\varepsilon > 0$.

\section{Simulation study, scenario 1: complete data}
To investigate the empirical behaviour of the proposed \textit{lqmHMM} in the presence of time-constant and time-varying sources of unobserved heterogeneity, we have implemented a large scale simulation study. Different experimental scenarios have been taken into consideration and the performances of the proposed approach have been compared with those of \textit{lqHMM}, obtained by setting $G = 1$ (this has been done to avoid extra-variability due to incoherence between the code we have developed ad that of \citealp{Farcomeni2012}).\\
Data have been generated from a two state mixed HMM ($m = 2)$ with initial/transition probabilities given by 
\begin{align}
\boldsymbol \delta = \left(0.7, 0.3 \right) \quad  \text{and} \quad  \textbf Q = \left(
\begin{matrix}
0.8 & 0.2 \\
0.2 & 0.8
\end{matrix} \right).
\end{align}
The following longitudinal regression model holds for the $h$-th state of the Markov chain:
\begin{equation}\label{cap5_simul_model}
Y_{it} = \alpha_{h} + (b_i +\beta_{1})\,  x_{it1} + \beta_{2} \, x_{it2} + \varepsilon_{it},
\end{equation}
where $x_{it1} \sim \text{Unif}[-10,10]$, $\tbf x_{it2} = \tbf x_{i2} \sim \text{Bin}(1, 0.5)$, $\beta_1 = 2$ and $\beta_{2}=-0.8$. Different probability distributions have been considered to generate the measurement error, $\varepsilon_{it}$, that is a standard Gaussian, and a Chi-square distribution with $\nu = 2$ degrees of freedom, to allow for skew data. 
The time-constant random slopes $b_i$ capture individual departures from the marginal effect $\beta_1$, and represent IID draws from a standard Gaussian ($\sigma_b = 1$) or a Student $t_3$ ($\sigma_b = \sqrt 3$) distribution. State-dependent intercepts have been set to $\alpha_1 = 100$ and $\alpha_2 = 110$.
The proposed simulation design follows the one discussed by \cite{LiuBottai2009} with a twofold aim. First, different distributions for the time-constant random parameters have been considered to study the robustness of the NPML approach when we move far from Gaussianity; second, different choices for the measurement error allow to study the performance of the proposed model in the presence of asymmetric errors. 
\\
Comparing results obtained when fitting the \textit{lqHMM} with those from the \textit{lqmHMM}, we expect a reduced efficiency of $\hat \beta_1$ in the former case, due to the extra-variability associated with $b_i$ which is not taken into consideration by this model specification; also, we expect this will produce some bias in the state-dependent intercepts and the associated latent Markovian structure.
\\
For each scenario we have generated $B = 250$ samples with varying sample sizes $(n = 100, 200)$ and lengths of longitudinal sequences $(T = 5, 10)$. Model parameters have been estimated for $\tau = (0.25, 0.50, 0.75)$. While we consider balanced designs only, the algorithm can be readily applied to unbalanced designs as well. For each simulated dataset, we have estimated a \textit{lmHMM} and a \emph{lqmHMM} with $m = 2$ hidden states; for the \emph{lqmHMM} specification, we have considered a varying number of mixture components ($G = 1, ... , 15$) and retained the model with the best BIC value. To evaluate the performances of the analysed models, we have computed the average bias and root mean square error (RMSE) for each model parameter. Simulation results for the parameters in the longitudinal model are reported in Tables \ref{simul_q25}-\ref{simul_q75}.

\begin{center}
Tables \ref{simul_q25}-\ref{simul_q75} ABOUT HERE
\end{center}

We may notice that the precision of parameter estimates is higher at the center of the distribution when compared to other quantiles both for \textit{lqHMM} and \textit{lqmHMM}. Such differences are due to the reduced amount of information at the tails and tend to decrease with increasing sample sizes and number of measurement occasions. 
\textbf{This result is particularly evident for the state-dependent intercepts. Indeed, based on the simulation scheme we have implemented, intercepts are directly related to the estimated quantiles, while the other parameters are constant when moving from $\tau = 0.25$ to $\tau = 0.75$. The precision of  $\hat \alpha_1$ and $\hat \alpha_2$ is, therefore, generally higher when estimating the center of the distribution than when focusing on the tails.} \\
The results we have obtained for $\tau = 0.25$ are worth of some discussion. Given the values used for $\bs \alpha, \bs \delta$ and $\tbf Q$, the first quartile of the conditional distribution mainly entails units in the first state of the Markov chain, with intercept $\alpha_1$. Therefore, $\alpha_2$ is always estimated with a lower precision than $\alpha_1$ due to a sparse (imprecise) information, but for the case $b_i \sim t_3$ when fitting the \textit{lqHMM} as $\alpha_2$ is less influenced by the (heavy) left tail of random effect distribution (not accounted for in \textit{lqHMM}).
{However, increasing the number of repeated measurements, $T$, we have a higher chance to observe transitions towards the second state and the quality of results clearly improves when fitting the \textit{lqmHMM}; bias and higher variability seem to persist when fitting \textit{lqHMM}. 
When comparing the two model approaches, it is clear that estimating a \textit{lqHMM} when both time-constant and time-varying sources of unobserved heterogeneity affect the longitudinal responses returns in a lower precision of all model parameters. A higher bias is observed for the state-dependent intercepts: the extra-variability not taken into consideration by the \textit{lqHMM} formulation is mostly captured by the latent Markov chain. }
However, given that $x_{i2}$ is time-constant, we may think that those sources of unobserved heterogeneity that do not evolve over time may also produce a high variability in the estimates for $\hat \beta_2$. RMSEs associated to $\hat \beta_2$ are generally higher (compared with those for $\hat \beta_1$) also when 
fitting the \textit{lqmHMM} possibly because of some aliasing between such a parameter and the discrete mixture-component locations.
The impact of this issue seems to reduce with increasing $T$ for \textit{lqmHMM}, while remain persistent in \textit{lqHMM}.
\\
Looking at the estimated variance of the time-constant, individual-specific, random parameter $b_i$ ($\sigma_b$), we may observe a reduced RMSE when this parameter is drawn from a standard Gaussian rather than from a $t_3$ distribution, which suggests that some extra-variability has been ruled out by the quantile regression. In fact, the estimated $\sigma_b$ is always lower than the true one ($\sigma_b = \sqrt 3$).
\\
To conclude, by looking at results reported in Tables \ref{simul_q25}-\ref{simul_q75}, it can be notice that estimates obtained by the \textit{lqmHMM} show a clear and consistent path, with precision increasing with $T$ and $n$, regardless of the measurement error distribution. As regards the random parameter distribution, some extra-variability in the parameter estimates is generally observed, especially for low sample sizes and length of the study.
On the other hand, the estimates obtained under the \textit{lqHMM} formulation are persistently biased and result somewhat influenced by the measurement error distribution and, more substantially, by the random parameter distribution.

\clearpage
 
\section*{Tables}

\begin{table}[!ht]
\centering
\caption{Simulation study for $\tau = 0.25$. Bias and RMSE for longitudinal parameter estimates under the \textit{lqHMM} and \textit{lqmHMM} formulation.} 
\scalebox{0.90}{

\begin{tabular}{llrrrrrrrrrrrrrrrrrrrrrrrrr}
 \toprule
  	
  &	& \multicolumn{2}{c}{$b_i \sim N(0,1), \epsilon_{it} \sim N(0,1)$} &
  	\multicolumn{2}{c}{$b_i \sim N, \epsilon_{it} \sim \chi_2^2$} &
  	\multicolumn{2}{c}{$b_i \sim t_3, \epsilon_{it} \sim \chi_2^2$} \\[0.2cm]
  	
&& \multicolumn{1}{c}{\textit{lqHMM}} &  \multicolumn{1}{c}{\textit{lqmHMM}}	& \multicolumn{1}{c}{\textit{lqHMM}} & \multicolumn{1}{c}{\textit{lqmHMM}}& \multicolumn{1}{c}{\textit{lqHMM}} &\multicolumn{1}{c}{\textit{lqmHMM}}\\
\hline
\vspace{-0.5cm}
\multirow{14}{*}{\rotatebox{90}{$n = 100$}}\\

&&\multicolumn{7}{l}{$\mathtt{T = 5}$}\\	

&		$\alpha_1$	&	-4.970	(6.25)	&	-0.229	(0.42)	&	1.224	(1.71)	&	0.167	(0.36)	&	-8.909	(13.53)	&	0.012	(0.71)	\\
&		$\alpha_2$	&	-6.784	(7.73)	&	-2.220	(4.34)	&	1.535	(1.82)	&	0.270	(0.47)	&	-7.330	(8.10)	&	-3.231	(5.36)	\\
&		$\beta_1$	&	0.015	(0.12)	&	0.008	(0.10)	&	0.014	(0.12)	&	0.008	(0.10)	&	0.010	(0.16)	&	0.020	(0.16)	\\
&		$\beta_2$	&	0.290	(1.68)	&	0.058	(0.29)	&	-0.168	(1.08)	&	-0.112	(0.42)	&	0.048	(1.51)	&	0.113	(0.68)	\\
&		$\sigma_b$	&			&	-0.100	(0.15)	&			&	-0.107	(0.14)	&			&	-0.340	(0.43)	\\

	\cline{2-10}							
					
													&	&\multicolumn{7}{l}{$\mathtt{T = 10}$}	\\

&		$\alpha_1$	&	-3.897	(5.16)	&	-0.176	(0.22)	&	-1.949	(3.13)	&	-0.007	(0.10)	&	-9.482	(13.12)	&	-0.069	(0.13)	\\
&		$\alpha_2$	&	-4.497	(5.99)	&	-0.246	(1.00)	&	-1.825	(3.31)	&	0.002	(0.10)	&	-5.729	(7.03)	&	-0.049	(0.12)	\\
&		$\beta_1$	&	0.008	(0.12)	&	0.007	(0.10)	&	0.004	(0.12)	&	0.003	(0.10)	&	0.016	(0.15)	&	0.022	(0.17)	\\
&		$\beta_2$	&	-0.102	(1.93)	&	0.041	(0.14)	&	0.014	(1.28)	&	0.016	(0.11)	&	0.217	(1.64)	&	0.045	(0.12)	\\
&		$\sigma_b$	&			&	-0.068	(0.11)	&			&	-0.070	(0.10)	&			&	-0.318	(0.42)	\\

\hline
\hline
\vspace{-0.5cm}

\multirow{15}{*}{\rotatebox{90}{$n = 200$}}\\

&&\multicolumn{7}{l}{$\mathtt{T = 5}$}\\	

&		$\alpha_1$	&	-6.105	(7.09)	&	-0.257	(0.35)	&	-2.679	(4.11)	&	-0.004	(0.10)	&	-12.994	(17.12)	&	0.018	(0.30)	\\
&		$\alpha_2$	&	-7.211	(8.10)	&	-0.914	(2.54)	&	-3.139	(4.88)	&	-0.007	(0.11)	&	-7.623	(8.38)	&	-1.599	(3.76)	\\
&		$\beta_1$	&	0.007	(0.10)	&	0.000	(0.07)	&	0.005	(0.10)	&	0.003	(0.07)	&	0.006	(0.12)	&	0.009	(0.12)	\\
&		$\beta_2$	&	0.032	(0.90)	&	0.030	(0.18)	&	0.046	(0.92)	&	0.015	(0.11)	&	-0.094	(0.71)	&	0.026	(0.20)	\\
&		$\sigma_b$	&			&	-0.091	(0.12)	&			&	-0.067	(0.09)	&			&	-0.266	(0.42)	\\

	\cline{2-10}							
					
													&&\multicolumn{7}{l}{$\mathtt{T = 10}$}	\\				
													
&		$\alpha_1$	&	-3.862	(5.10)	&	-0.108	(0.13)	&	-1.285	(1.80)	&	0.014	(0.06)	&	-12.395	(16.85)	&	-0.030	(0.07)	\\
&		$\alpha_2$	&	-3.995	(5.67)	&	-0.096	(0.12)	&	-0.918	(1.76)	&	0.018	(0.07)	&	-6.190	(7.41)	&	-0.006	(0.07)	\\
&		$\beta_1$	&	0.003	(0.09)	&	0.001	(0.07)	&	0.011	(0.09)	&	0.000	(0.07)	&	-0.001	(0.10)	&	0.007	(0.12)	\\
&		$\beta_2$	&	0.129	(1.54)	&	0.007	(0.08)	&	0.031	(0.81)	&	0.003	(0.07)	&	0.048	(1.57)	&	0.009	(0.08)	\\
&		$\sigma_b$	&			&	-0.047	(0.07)	&			&	-0.054	(0.08)	&			&	-0.272	(0.41)	\\

 \bottomrule
\bottomrule
\end{tabular}\label{simul_q25}

}

\end{table}

\begin{table}[!ht]
\centering
\caption{Simulation study for $\tau = 0.50$. Bias and RMSE for longitudinal parameter estimates under the \textit{lqHMM} and \textit{lqmHMM} formulation.} 
\scalebox{0.90}{

\begin{tabular}{llrrrrrrrrrrrrrrrrrrrrrrrrr}
 \toprule
  	
   &	& \multicolumn{2}{c}{$b_i \sim N(0,1), \epsilon_{it} \sim N(0,1)$} &
  	\multicolumn{2}{c}{$b_i \sim N, \epsilon_{it} \sim \chi_2^2$} &
  	\multicolumn{2}{c}{$b_i \sim t_3, \epsilon_{it} \sim \chi_2^2$} \\[0.2cm]
  	
&& \multicolumn{1}{c}{\textit{lqHMM}} &  \multicolumn{1}{c}{\textit{lqmHMM}}	& \multicolumn{1}{c}{\textit{lqHMM}} & \multicolumn{1}{c}{\textit{lqmHMM}}& \multicolumn{1}{c}{\textit{lqHMM}} &\multicolumn{1}{c}{\textit{lqmHMM}}\\
\hline
\vspace{-0.5cm}
\multirow{14}{*}{\rotatebox{90}{$n = 100$}}\\

&&\multicolumn{7}{l}{$\mathtt{T = 5}$}\\	

&		$\alpha_1$	&	-0.131	(0.36)	&	-0.028	(0.12)	&	0.150	(0.48)	&	0.145	(0.26)	&	0.366	(0.62)	&	0.200	(0.32)	\\
&		$\alpha_2$	&	-0.003	(0.43)	&	-0.002	(0.13)	&	0.196	(0.60)	&	0.133	(0.27)	&	0.064	(0.63)	&	0.134	(0.28)	\\
&		$\beta_1$	&	0.018	(0.12)	&	0.012	(0.10)	&	0.017	(0.13)	&	0.008	(0.10)	&	0.003	(0.15)	&	0.012	(0.16)	\\
&		$\beta_2$	&	0.073	(0.43)	&	0.027	(0.15)	&	0.086	(0.57)	&	0.011	(0.26)	&	0.023	(0.64)	&	-0.034	(0.29)	\\
&		$\sigma_b$	&			&	-0.104	(0.13)	&			&	-0.109	(0.14)	&			&	-0.459	(0.51)	\\

	\cline{2-10}							
					
													&	&\multicolumn{7}{l}{$\mathtt{T = 10}$}	\\

&		$\alpha_1$	&	-0.082	(0.31)	&	-0.016	(0.09)	&	0.202	(0.40)	&	0.116	(0.18)	&	0.359	(0.54)	&	0.105	(0.17)	\\
&		$\alpha_2$	&	0.031	(0.28)	&	-0.003	(0.10)	&	0.259	(0.44)	&	0.100	(0.19)	&	0.148	(0.47)	&	0.135	(0.20)	\\
&		$\beta_1$	&	0.013	(0.11)	&	0.002	(0.07)	&	0.011	(0.11)	&	0.001	(0.07)	&	0.023	(0.14)	&	0.022	(0.16)	\\
&		$\beta_2$	&	0.021	(0.33)	&	0.018	(0.10)	&	-0.007	(0.42)	&	0.016	(0.16)	&	0.019	(0.44)	&	-0.004	(0.17)	\\
&		$\sigma_b$	&			&	-0.075	(0.09)	&			&	-0.074	(0.10)	&			&	-0.402	(0.45)	\\
\hline
\hline
\vspace{-0.5cm}

\multirow{15}{*}{\rotatebox{90}{$n = 200$}}\\

&&\multicolumn{7}{l}{$\mathtt{T = 5}$}\\	

&		$\alpha_1$	&	-0.076	(0.25)	&	0.002	(0.06)	&	0.186	(0.36)	&	0.089	(0.12)	&	0.383	(0.55)	&	0.160	(0.22)	\\
&		$\alpha_2$	&	0.062	(0.31)	&	0.003	(0.06)	&	0.235	(0.46)	&	0.091	(0.13)	&	0.099	(0.46)	&	0.125	(0.21)	\\
&		$\beta_1$	&	0.006	(0.09)	&	0.002	(0.07)	&	0.009	(0.09)	&	0.001	(0.07)	&	-0.001	(0.11)	&	0.010	(0.12)	\\
&		$\beta_2$	&	0.024	(0.32)	&	-0.007	(0.06)	&	0.013	(0.40)	&	-0.004	(0.10)	&	-0.007	(0.45)	&	-0.005	(0.17)	\\
&		$\sigma_b$	&			&	-0.056	(0.08)	&			&	-0.054	(0.08)	&			&	-0.388	(0.43)	\\

	\cline{2-10}							
					
													&&\multicolumn{7}{l}{$\mathtt{T = 10}$}	\\				
											&		$\alpha_1$	&	-0.080	(0.21)	&	0.002	(0.06)	&	0.192	(0.32)	&	0.116	(0.18)	&	0.335	(0.43)	&	0.085	(0.12)	\\
&		$\alpha_2$	&	0.055	(0.20)	&	0.003	(0.06)	&	0.261	(0.39)	&	0.100	(0.19)	&	0.123	(0.33)	&	0.106	(0.14)	\\
&		$\beta_1$	&	0.009	(0.09)	&	0.002	(0.07)	&	0.013	(0.08)	&	0.001	(0.07)	&	0.005	(0.10)	&	0.011	(0.12)	\\
&		$\beta_2$	&	0.009	(0.22)	&	-0.007	(0.06)	&	-0.008	(0.31)	&	0.016	(0.16)	&	0.026	(0.32)	&	0.011	(0.12)	\\
&		$\sigma_b$	&			&	-0.056	(0.08)	&			&	-0.074	(0.10)	&			&	-0.328	(0.38)	\\

 \bottomrule
\bottomrule
\end{tabular}\label{simul_q50}

}

\end{table}

\begin{table}[!ht]
\centering
\caption{Simulation study for $\tau = 0.75$. Bias and RMSE for longitudinal parameter estimates under the \textit{lqHMM} and \textit{lqmHMM} formulation.} 
\scalebox{0.90}{

\begin{tabular}{llrrrrrrrrrrrrrrrrrrrrrrrrr}
 \toprule
  	
  &	& \multicolumn{2}{c}{$b_i \sim N(0,1), \epsilon_{it} \sim N(0,1)$} &
  	\multicolumn{2}{c}{$b_i \sim N, \epsilon_{it} \sim \chi_2^2$} &
  	\multicolumn{2}{c}{$b_i \sim t_3, \epsilon_{it} \sim \chi_2^2$} \\[0.2cm]
  	
&& \multicolumn{1}{c}{\textit{lqHMM}} &  \multicolumn{1}{c}{\textit{lqmHMM}}	& \multicolumn{1}{c}{\textit{lqHMM}} & \multicolumn{1}{c}{\textit{lqmHMM}}& \multicolumn{1}{c}{\textit{lqHMM}} &\multicolumn{1}{c}{\textit{lqmHMM}}\\
\hline
\vspace{-0.5cm}
\multirow{14}{*}{\rotatebox{90}{$n = 100$}}\\

&&\multicolumn{7}{l}{$\mathtt{T = 5}$}\\	

&		$\alpha_1$	&	1.109	(1.45)	&	0.170	(0.23)	&	1.741	(2.41)	&	0.158	(0.39)	&	3.015	(3.92)	&	0.389	(0.61)	\\
&		$\alpha_2$	&	1.632	(1.90)	&	0.213	(0.28)	&	2.045	(2.54)	&	0.343	(0.53)	&	3.792	(5.95)	&	0.384	(0.59)	\\
&		$\beta_1$	&	0.008	(0.12)	&	0.007	(0.10)	&	0.023	(0.13)	&	0.022	(0.10)	&	0.003	(0.16)	&	0.014	(0.16)	\\
&		$\beta_2$	&	-0.107	(0.95)	&	-0.045	(0.20)	&	-0.287	(1.36)	&	-0.174	(0.46)	&	-0.116	(1.91)	&	-0.151	(0.73)	\\
&		$\sigma_b$	&			&	-0.108	(0.14)	&			&	-0.102	(0.13)	&			&	-0.375	(0.48)	\\

	\cline{2-10}							
					
													&	&\multicolumn{7}{l}{$\mathtt{T = 10}$}	\\				

&		$\alpha_1$	&	1.020	(1.16)	&	0.120	(0.16)	&	1.388	(1.92)	&	-0.056	(0.21)	&	3.907	(5.12)	&	0.023	(0.27)	\\
&		$\alpha_2$	&	1.488	(1.58)	&	0.129	(0.16)	&	1.869	(2.28)	&	0.111	(0.25)	&	7.327	(16.36)	&	0.229	(0.36)	\\
&		$\beta_1$	&	0.010	(0.12)	&	0.006	(0.09)	&	0.006	(0.12)	&	0.003	(0.10)	&	0.017	(0.15)	&	0.022	(0.16)	\\
&		$\beta_2$	&	-0.104	(0.70)	&	-0.016	(0.12)	&	-0.122	(1.03)	&	-0.064	(0.27)	&	0.196	(2.04)	&	-0.041	(0.29)	\\
&		$\sigma_b$	&			&	-0.079	(0.11)	&			&	-0.062	(0.10)	&			&	-0.320	(0.40)	\\

\hline
\hline
\vspace{-0.5cm}

\multirow{15}{*}{\rotatebox{90}{$n = 200$}}\\

&&\multicolumn{7}{l}{$\mathtt{T = 5}$}\\	

&		$\alpha_1$	&	0.950	(1.04)	&	0.104	(0.14)	&	1.161	(1.32)	&	0.003	(0.23)	&	2.593	(3.39)	&	0.181	(0.35)	\\
&		$\alpha_2$	&	1.507	(1.59)	&	0.136	(0.18)	&	1.689	(1.79)	&	0.162	(0.31)	&	3.843	(6.76)	&	0.274	(0.43)	\\
&		$\beta_1$	&	0.010	(0.10)	&	0.006	(0.07)	&	0.009	(0.09)	&	0.003	(0.08)	&	0.005	(0.12)	&	0.015	(0.12)	\\
&		$\beta_2$	&	-0.057	(0.53)	&	0.003	(0.11)	&	-0.085	(0.69)	&	-0.062	(0.27)	&	-0.213	(1.66)	&	-0.096	(0.36)	\\
&		$\sigma_b$	&			&	-0.083	(0.10)	&			&	-0.069	(0.09)	&			&	-0.310	(0.39)	\\

	\cline{2-10}							
					
													&&\multicolumn{7}{l}{$\mathtt{T = 10}$}	\\				
													
&		$\alpha_1$	&	0.931	(0.99)	&	0.094	(0.11)	&	1.183	(1.48)	&	-0.129	(0.19)	&	3.393	(4.63)	&	-0.062	(0.17)	\\
&		$\alpha_2$	&	1.511	(1.55)	&	0.111	(0.13)	&	1.732	(1.93)	&	0.076	(0.18)	&	5.919	(10.44)	&	0.142	(0.22)	\\
&		$\beta_1$	&	0.010	(0.09)	&	0.002	(0.07)	&	0.005	(0.08)	&	0.003	(0.07)	&	0.009	(0.10)	&	0.015	(0.12)	\\
&		$\beta_2$	&	-0.073	(0.45)	&	-0.016	(0.08)	&	-0.075	(0.60)	&	-0.027	(0.18)	&	0.090	(1.44)	&	0.004	(0.19)	\\
&		$\sigma_b$	&			&	-0.059	(0.08)	&			&	-0.048	(0.07)	&			&	-0.257	(0.42)	\\

 \bottomrule
\bottomrule
\end{tabular}\label{simul_q75}

}

\end{table}

\clearpage

\bibliographystyle{plainnat}
\bibliography{biblio}